\documentclass{nature}

\usepackage{amssymb}
\usepackage{amsmath}
\usepackage{amsthm}
\usepackage{graphicx}
\usepackage{lineno}
\usepackage[colorlinks=true,linkcolor=blue,citecolor=blue,pdfauthor={ },pdftitle={ },pdfsubject={ },pdfkeywords={ }]{hyperref}
\graphicspath{{Figures/}}

\newcommand{\fish}[0]{F_\mathrm{Q}}
\title{Probe optimization for quantum metrology via closed-loop learning control}

\author{Xiaodong Yang$^{1}$, Jayne Thompson$^2$, Ze Wu$^{1}$, Mile Gu$^{3,4,2,*}$, Xinhua Peng$^{1,5,6,*}$, \& Jiangfeng Du$^{1,5,6}$}

\begin{document}
\maketitle

\begin{affiliations}
 \item CAS Key Laboratory of Microscale Magnetic Resonance and Department of Modern Physics, University of Science and Technology of China, Hefei, Anhui 230026, China
 \item Centre for Quantum Technologies, National University of Singapore, 3 Science Drive 2, Singapore 117543
 \item Nanyang Quantum Hub, School of Physical and Mathematical Sciences Nanyang Technological University Singapore, Singapore
 \item Complexity Institute, Nanyang Technological University, Singapore and Centre for Quantum Technologies, National University of Singapore, Singapore
 \item Hefei National Laboratory for Physical Sciences at the Microscale, University of Science and Technology of China, Hefei 230026, China
 \item Synergetic Innovation Centre of Quantum Information $\&$ Quantum Physics, University of Science and Technology of China, Hefei, Anhui 230026, China
\end{affiliations}

\begin{abstract}
Experimentally achieving the precision that standard quantum metrology schemes promise is always challenging. Recently, additional controls were applied to design feasible quantum metrology schemes. However, these approaches generally does not consider ease of implementation, raising technological barriers impeding its realization. In this paper, we circumvent this problem by applying closed-loop learning control to propose a practical controlled sequential scheme for quantum metrology.  Purity loss of the probe state, which relates to quantum Fisher information, is measured efficiently as the fitness to guide the learning loop. We confirm its feasibility and certain superiorities over standard quantum metrology schemes by numerical analysis and proof-of-principle experiments in a nuclear magnetic resonance (NMR) system.
\end{abstract}


\section*{INTRODUCTION}
Much of quantitative science deals with measuring a certain parameter, say $\phi$, of a physical process precisely. Typically, this involves subjecting suitably engineered probe states to the physical process, and using measurement readout to recover an estimate of $\phi$. The central limit theorem states that repeated applications of this procedure can improve our estimate, such that the resulting standard error scales as ${1}/{\sqrt{N}}$ in the number of particles $N$. Remarkably, quantum technologies allow us to surpass this standard limit. By using suitably entangled probes, we can reach the \emph{Heisenberg limit} -- suppressing $\Delta \phi$ such that it scales as $1/N$~\cite{GV06,GI11,GT14}. This quadratic scaling advantage can drastically reduce the resources required for precision measurement, and continues to catalyze rapid developments in the field of quantum metrology~\cite{KLS14,LJL15,YF15,YHD16,HWJ19,OLO12,HBD07,BGN15,BBD16,HS10,PGS17,PJ17}.

While quantum metrology is well understood at the theoretical level, its physical application to large-scale quantum systems faces significance challenges \cite{PCL12,JKF09,SJK10,MMJ04,RPP07,JJM11,EDR11,EBMM11}. Consider the iconic task of estimating the phase $\phi$ of some unitary process $U_{\phi} =   e^{-\mathrm{i} \mathcal{H} \phi}$. In this setting, theory tells us how to determine the optimal quantum probe $\rho$ for any possible Hamiltonian $\mathcal{H}$. However, when $\mathcal{H}$ acts on a many-body system, this optimal probe is typically a complex, entangled many-body state. Engineering this probe is often non-trivial, especially in the advent of limited access to the physical operations used to synthesize such probes \cite{PCL12}. Meanwhile, many realistic means of initializing such probes involve applying a sequence of controls, whose operational effects are not fully characterized \cite{JJM11,WCL16,LKS05,JD13}. These issues are further exacerbated by the exponentially growing size of the Hilbert space -- making direct implementation of complex metrological schemes extremely challenging.

Here, we propose a closed-loop learning protocol that circumvents these issues. The resulting protocol has the following desirable features: (a) It does not require us to analytically solve for the optimal probe -- nor possess prior-knowledge of how this probe can be synthesized from available physical controls. These aspects are optimized through the learning process. (b) It does not require us to know the precise effects of these physical controls, nor implement tomography on the resulting quantum probes. (c) It does not require any computation involving matrix representations of $\mathcal{H}$, avoiding the curse of dimensionality. These features combined allow a versatile procedure for finding improved metrology protocols, ideal for complex many-body settings. We demonstrate a proof-of-principle experiment using nuclear magnetic resonance (NMR), illustrating the viability of this approach with present-day quantum technology.

\section*{RESULTS}
\textbf{Framework of closed-loop learning assisted quantum metrology} \\
Here, we consider estimating the phase $\phi$ of a general $N$-body unitary process $U_{\phi} =  e^{-\mathrm{i}\mathcal{H}\phi}$. Each metrology protocol begins with a probe initialized in some easily prepared state $\rho_0$ on $N$ probes -- typically a product state where each probe is initialized in some default state $|0\rangle$. The goal then is to implement a control -- a sequence of physical operations that we apply to the $N$-probe system, transforming $\rho_0$ to some candidate entangled state $\rho_C$. By acting  $U_{\phi}$ on each probe, we end up with $\rho_{\phi}$, that encodes information regarding $\phi$ (See Fig. \ref{scheme}\textbf{a}). The efficacy of each candidate probe $\rho_C$ is typically quantified by the quantum Fisher information $\fish$ \cite{GI11}. The rationale is that after repeating this process through $\nu$ independent runs , the standard error to which we can estimate $\phi$ is bounded below by $1/\sqrt{\nu \fish}$. This lower bound is tight, and can always be saturated using an ideal measurement scheme.

The particular benefit of quantum metrology is that use of suitably entangled $\rho_C$ enables one to reduced uncertainty of $\phi$ much more quickly than conventional strategies. One iconic case, for example, is when $U_{\phi}$ corresponds applying an identical unitary process $e^{-\mathrm{i} \mathcal{H} \phi}$ to each individual probe. In such scenarios, use of non-entangled $\rho_C$ results in $\fish$ that scales linearly with $N$, such that sensing $\phi$ to some desired $\Delta$ requires $N > \mathrm{O}(1/\Delta^2)$ probes. In contrast, if $\rho_C$ is appropriately entangled, $\fish$ that scales as $N^2$, enabling the Heisenberg limit scaling of $N > \mathrm{O}(1/\Delta)$. The goal of quantum metrology can thus be split into two distinct tasks.
\begin{itemize}
    \item[1.] Determine the control sequence $C$ that synthesizes some near optimal state $\rho_C$ whose corresponding quantum Fisher information $\fish(\rho_C)$ is made as large as possible.
    \item[2.] Use the control sequence to synthesize $\rho_C$, which can then be injected as input to $U_{\phi}$ for purposes of estimating $\phi$.
\end{itemize}
Here, our primary is the first task, with understanding that our resulting control sequences can be used to synthesize the appropriate states to perform metrology. This is highly non-trivial for general $\mathcal{H}$. Notably, the dimensions of $\mathcal{H}$ grow exponentially with $N$, making analytical methods for finding the optimal $\rho_C$ computationally intractable. Meanwhile, $C$ is described by an ordered list of readily accessible elementary operations (e.g., pulse sequences). Inferring how these can be chained together to generate a given $\rho_C$ is generally highly non-trivial, especially when the exact physical effect of each elementary operation on the probe state is not known. Typical means of optimizing $\fish(\rho_C)$ are further hampered by difficulty in evaluating the efficacy of a candidate control sequence $C$. Given $\rho_{\phi}$, evaluation of the corresponding efficacy $\fish(\rho_C)$ involves an optimization over all possible measurement bases -- a task whose complexity also scales exponentially with system size.

In our protocol, we first tackle the difficulty in evaluating efficacy by using relations between quantum Fisher information and purity loss. Let $\rho_{\mathrm{avg}} = \int P_{x} \rho_{\phi + x} \mathrm{d}x $, where $P_x$ is some probability distribution with mean $0$ and standard deviation $\Delta x$. Meanwhile setting $\Delta\gamma(\Delta x) = \text{Tr}(\rho_C^2) - \text{Tr}(\rho_{\mathrm{avg}}^2)$. Then recent results \cite{MLJM16} established that in the limit where $\Delta x \ll 1$, the quantum Fisher information $\fish(\rho_C)$  with respect to $\phi$ satisfies
\begin{eqnarray}\label{relationship}
F_\mathrm{Q}(\rho_C)\ge 2\frac{\Delta \gamma (\Delta x)}{(\Delta x)^2} := F_\mathrm{Q}^{\mathrm{L}}.
\end{eqnarray}

 Physically $\rho_{\mathrm{avg}}$ represents the resulting ensemble state when the aforementioned metrology procedure is applied to a unitary $U_{\phi}$ such that $\phi$ undergoes stochastic fluctuations of magnitude $\Delta x$. Thus $\Delta\gamma(\Delta x)$ captures the purity loss of the resultant state induced by these fluctuations. Eq. \ref{relationship} then states that the efficacy of a metrology protocol is bounded below by the rate in which its output state loses purity when subject to stochastic noise in the parameter we are trying to sense.
Therefore, we can effectively use $\fish^{\mathrm L}$ as a proxy for the efficacy of a probe.

The advantage is that purity loss is far more amendable to direct measurement than quantum Fisher information~\cite{ACDM02}. To evaluate the efficacy of a candidate $C$, we apply two pairs of the control sequence in parallel to obtain two copies of $\rho_C$. The rate of purity loss of the resulting outputs when subject to stochastic noise on $\phi$ can then be experimentally measured by application of suitable controlled-SWAP gates -- coherently swapping output pairs controlled on an ancillary quantum mechanical degree of freedom (See Fig. \ref{scheme}\textbf{c}). We refer to this quantum algorithm as the \emph{quantum efficacy estimator}, which can now be coupled with a suitable closed-loop learning algorithm for automated discovery of increasingly effective control sequences for sensing $\phi$ (See Fig. \ref{scheme}\textbf{b}).

In practice, we can use many different learning algorithms, ranging from simple direct search algorithms \cite{LTT00} to more complex evolutionary algorithms \cite{ES15}. Here, we found the Nelder-Mead algorithm \cite{NM65} to be particular effective for our experiments. The entire learning process can then be summarised as follows: we begin by initializing a population of $n+1$ control sequences at random, and make use of the quantum efficacy estimator to sort them in the order of decreasing efficacy, denoted $\mathcal{C}^{(g)} = \{C_0^{(g)}, C_1^{(g)}, \ldots, C_{n+1}^{(g)}\}$, with $g = 0$ indicating the $0^{\mathrm{th}}$ iteration. Meanwhile $n$ is generally chosen to scale linearly with the number of actions (controls) we can apply in each particular time-step (e.g., if we have access to local rotations along $x$ and $y$ axis on $N$ qubits, then $n$ scales linearly with $N$). Once the initial population is set, the Nelder-Mead algorithm then stipulates a systematic method to generate a new candidate control through geometric considerations -- which replaces the worst performer $C_{n+1}^{(g)}$ to form the population in the next iteration, which is then again sorted by decreasing efficacy to obtain $\mathcal{C}^{(g+1)}$. The exact mechanics of this algorithms involve mapping each control sequence into a vertex in some suitable convex space, the details of which are found in Methods. At each iteration $g$, the control sequence with maximum purity loss is denoted as $C^{(g)}$. The procedure is then continued until some designated stopping condition, such as when the purity loss of $C^{(g)}$ becomes sufficiently stable over multiple rounds, or when a set number of iterations are reached. Once the stopping condition is hit, the $C^{(g)}$ with maximum efficacy is delivered as the recommended control sequence.

This optimization process scales as a polynomial with respect to the number of free parameters that specifies a candidate control sequence. This latter condition is typically true for realistic settings, where (1) we are typically limited to one and two-body interactions, such that the number of possible control sequences we have at any particular point in time scales at most scales quadratically with $N$, (2) we are reasonably restricted to the reachable states within polynomial resource, such that the amount of time it takes to synthesize them does not scale exponentially with $N$. The protocol also has a number of other key advantages. First, at no stage does it require tomography of candidate probe states, either before or after the action of $U_\phi$. Secondly, it does not require complete mathematical knowledge of how the control sequences act on the Hilbert space. Finally, the algorithm automatically accounts for potentially physical anomalies, such as drift Hamiltonians when optimizing the control sequences of synthesizing the probe state. Each of these tasks would typically require a classical computer an exponentially amount of time to address. Thus the learning procedure inherits all the advantages of closed-loop learning -- easily adaptable to diverse physical architectures \cite{DS13,BCM19,OJF09}.

\noindent
\textbf{Example of sensing with spin chains} \\
We illustrate these advantages numerically for a scenario with spin chains featuring unavoidable spin-spin interactions. Consider the case where we have access to $N$-qubit spin chains, and wish to use them to estimate the strength of some external magnetic field in the $z$ axis. This problem aligns with estimating the phase $\phi$ of an $N$-qubit unitary $U_\phi = e^{- \mathrm{i} \phi\sum\nolimits_{i = 1}^N I_z^i}$, where $I_z^i$ represents the angular momentum operator of the $i^{\mathrm{th}}$ spin. If non-entangled qubits are used, the achieved Fisher information will scale as $N$. In contrast, the use of appropriately entangled probes can lead a quantum Fisher information $\fish$ that scales as $N^2$, enabling us to achieve the Heisenberg limit. While theory would enable us to work out the optimal probe, the catch here is that our control on the spin system is limited. Firstly, the chain evolves naturally according to nearest-neighbor Ising coupling $\mathcal{H}_S = 2\pi J\sum\nolimits_{i = 2}^N {I_z^{i - 1}} I_z^i$ ($J$ is the coupling strength). Secondly, our control of the system is limited to a sequence of local $I_x$ and $I_y$ interactions, whose strength we can adjust $M$ times. What is then the optimal way to adjust our control fields? The question is non-trivial.

To apply our algorithm, we first formally describe $C$. Here, each time we adjust the control field, we have $M(2N+1)$ free parameters. (1) $2M$ parameters $B^i_{x}[m]$ and  $B^i_{y}[m]$ describing the strength of the spin angular momentum operators $I_x^i$ and $I_y^i$ for the $i^{\mathrm{th}}$ qubit for each $i = 1,2,\ldots, N$ and $m = 1,2,\ldots,M $, (2) $M$ parameters $\Delta t[m]$ describing the amount of time we should wait before adjusting the fields again. As such, each control sequence is described by $M(2N+1)$ parameters, $C = (B^i_{x}[m], B^i_{y}[m], \Delta t[m])$ where $m=1,2,..., M;i=1,2,...,N$. Application of this control sequence would then correspond to enacting the unitary $U_C = U_M \ldots U_2U_1$, where ${U_m} = {e^{ - \mathrm{i}\Delta t[m] 2\pi \{ J\sum_{i = 2}^N {I_z^{i - 1}I_z^i}  + \sum_{i = 1}^N {(B_x^i[m]I_x^i}  + B_y^i[m]I_y^i)\} }}$. The goal is to find some near optimal sequence, such that from some easily prepared initial probe $|{\Psi _{\mathrm i}} \rangle$ the resulting probe state $|{\Psi_{\mathrm f}} \rangle= U_C|{\Psi _{\mathrm i}} \rangle$ is near optimal for estimating $\phi$.

\begin{figure*}
  \centering
  \includegraphics[width=0.9\textwidth,height=0.55\textwidth]{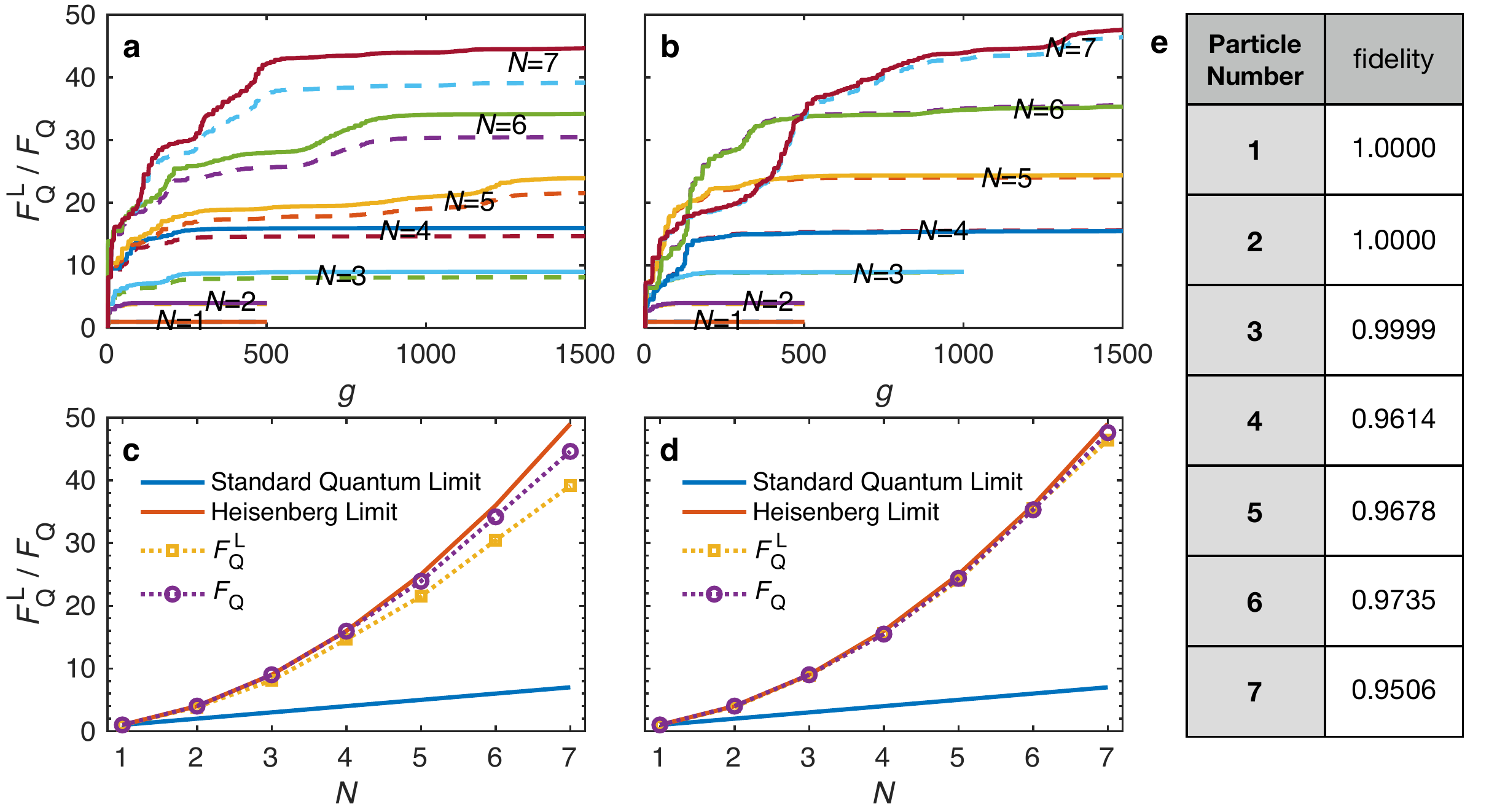}
  \caption{\textbf{Numerical simulations of the closed-loop learning algorithm when sensing with spin chains.} \textbf{a} and \textbf{b} each illustrate the performance of our closed-loop learning algorithms for the respective cases where $(\Delta x)^2 = 0.01$ and $(\Delta x)^2 = 0.001$. In both graphs, the horizontal axis denotes iteration number $g$. The solid lines represent optimal quantum Fisher information achieved at each iteration, while the dashed lines represent the bound stipulated by purity loss (i.e., $F^\mathrm{L}_\mathrm{Q}$). We see that both begin at low values at $g = 0$ as expected for random probes, and improve markedly during the learning process. \textbf{c} and \textbf{d} illustrate that the efficacy of the discovered probes approaches the Heisenberg limit, indicating their near optimality. Meanwhile setting $\Delta x$ to be smaller seemed to be marginally more advantageous, likely owing to the closer agreement between purity loss and quantum Fisher information in this regime. In \textbf{e}, we show a table of the fidelity between the quantum probe states generated and the closest theoretically optimal probe.}\label{simulation}
\end{figure*}

For sufficiently low $N$ (of up $7$), it is feasible to simulate our algorithm classically. In Fig. \ref{simulation}\textbf{a},\textbf{b}, we plot the maximum $\fish^{\mathrm L}$ and $\fish$ with respect to the particle number for two possible choices of $\Delta x$. Observe that the purity loss becomes a better proxy for quantum Fisher information when $\Delta x$ is reduced, in agreement with Eq. $\ref{relationship}$. Indeed, at $(\Delta x)^2 = 0.001$, the relationship is almost exact. As such, our learning protocols produce results within 1\% of the Heisenberg limit when using $(\Delta x)^2 = 0.001$.

To further verify the effectiveness of our algorithm, we compare the results of our algorithm with that of the theoretical optimal. In this specific case, theory indicates that the $N$-party entangled NOON states $|{\Psi _{\mathrm t}}\rangle  = (| 0 \rangle ^{\otimes N}+ e^{{\mathrm i} \theta}| 1 \rangle ^{\otimes N})/\sqrt{2}$ are the optimal probes - saturating the Heisenberg limit~\cite{HLW12}. Executing our closed-loop learning algorithm, we note that the learned optimal probe state $|\Psi_{\mathrm f}\rangle $ closely approximates these NOON states. In Fig. \ref{simulation}\textbf{c}, we list the fidelity $\langle \Psi_{\mathrm t}|\Psi_{\mathrm f} \rangle$ between ${|\Psi _{\mathrm f}\rangle}$ and theoretical optimal ${|\Psi _{\mathrm t}\rangle}$. For small qubit numbers, the agreement is complete (fidelity $=1$). While limitations in computational resources (in evaluating $\fish$ for example) do slowly degrade the fidelity as we increase particle number, there is still a match of over 0.95 when $N = 7$.

The algorithm, itself, however, is not designed to run purely on classical computers. Indeed, the computational costs to do so scale exponentially with $N$. Tracking the dynamics of controls, and resulting purity loss becomes quickly intractable. However, when the algorithm is executed on a quantum processor, such information does not need to be tracked. In particular, we do not need to know the mathematical descriptions of the controls, nor the strength of the internal spin-spin interactions.

\noindent
\textbf{Proof-of-principle experiment} \\
Our proof-of-principle experiment was conducted on a Bruker Avance III 400 MHz spectrometer using the sample Diethyl-fluoromalonate at room temperature. This three-qubit nuclear magnetic resonance processor (MNR) consists of three spins $^{13}\text{C}$, $^{1}\text{H}$ and $^{19}\text{F}$. Label these as qubits $1$, $2$ and $3$. This process enables us to engineer controlled-SWAP gates that coherently swaps between qubits $2$ and $3$, controlled on qubit $1$ (see Methods and Supplementary Note 3). Thus provided we can initialize both qubits $2$ and $3$ in a designated state $\rho$, we can experimentally measure the purity $\mathrm{Tr}(\rho^2)$~\cite{ACDM02}. This processor is thus capable of realising the quantum efficacy estimator for single qubit probes.

We illustrate the use of this device to estimate $\phi$, encoded within the single qubit unitary $U_{\phi}=e^{-\mathrm{i} I_z \phi}$. Here the probe state is a single spin,  which we can rotate along $x$ and $y$ directions. Assume $M$ total pulse segments, each candidate control sequence is now described by $2M$ free parameters, $C = (B_x [1], B_y[1],  B_x [2], B_y[2], \ldots  B_x [M], B_y[M])$. The resulting propagator could be expressed as $U_C = U_M\ldots U_2U_1$ with
\begin{eqnarray}
U_m = \exp \{-\mathrm{i} ( B_x[m] I_x + B_y[m] I_y ) \Delta t \}.
\end{eqnarray}
Note that we have omitted the $\Delta t [m]$ which was present in numerical simulation for the general $N$ case, as the lack of a drift Hamiltonian makes this unnecessary. Our goal is then to find a control sequence $C$ that such that $U_C |0\rangle$ has maximal quantum Fisher information with respect to $\phi$.

We implement our closed-loop learning algorithm with a population of $n = 7$, and $M = 3$ pulse segments. Each pulse sequence was set to $T=M\tau=30~\rm{\mu s}$. The key difference here from numerics is that the efficacy is now evaluated directly  using our NMR processor. For a particular candidate control sequence $C$, we first initialize each of qubits 2 and 3 of our processor into the state $|0\rangle$. The control sequence $C$ is then applied to both qubits, setting them each to some resulting candidate probe state $\rho_C$. Application of the controlled-SWAP circuit then enables estimation of $\gamma_C = \text{Tr}(\mathrm{\rho^2_C})$ (see Fig. \ref{scheme}\textbf{c}).

Determination of $\gamma_{\mathrm{avg}} =\text{Tr}(\rho^2_{\mathrm{avg}})$, requires us to simulate the effects of applying $U_{\phi + X}$, where $X$ is Gaussian distribution with standard deviation $\Delta x$. This is a little more complex in the NMR regime, but can be done using a variation of stratified sampling (see Methods). Once done, we can then directly evaluate the efficacy estimator $\Delta \gamma = \gamma_C - \gamma_{\mathrm{avg}}$ (see Fig. \ref{scheme}\textbf{b}). Thus our NMR processor is able to function as an effective quantum efficacy estimator.

This gives us all the tools in place for a quantum assisted closed-loop learning algorithm. To begin, we generated a random selection of $7$ control sequences, denoted as $\mathcal{C}^{(0)}$. By evaluating their efficacy using the NMR processor, and feeding results into the Nelder-Mead algorithm, we can systematically produce subsequent populations $\mathcal{C}^{(1)}$, $\mathcal{C}^{(2)}$, $\ldots $. We emphasize that the entire procedure was fully automated, such that this procedure can proceed ad-infinitum without intervention till stopping conditions are met.

\begin{figure*}
  \centering
  \includegraphics[width=0.98\textwidth,height=0.35\textwidth]{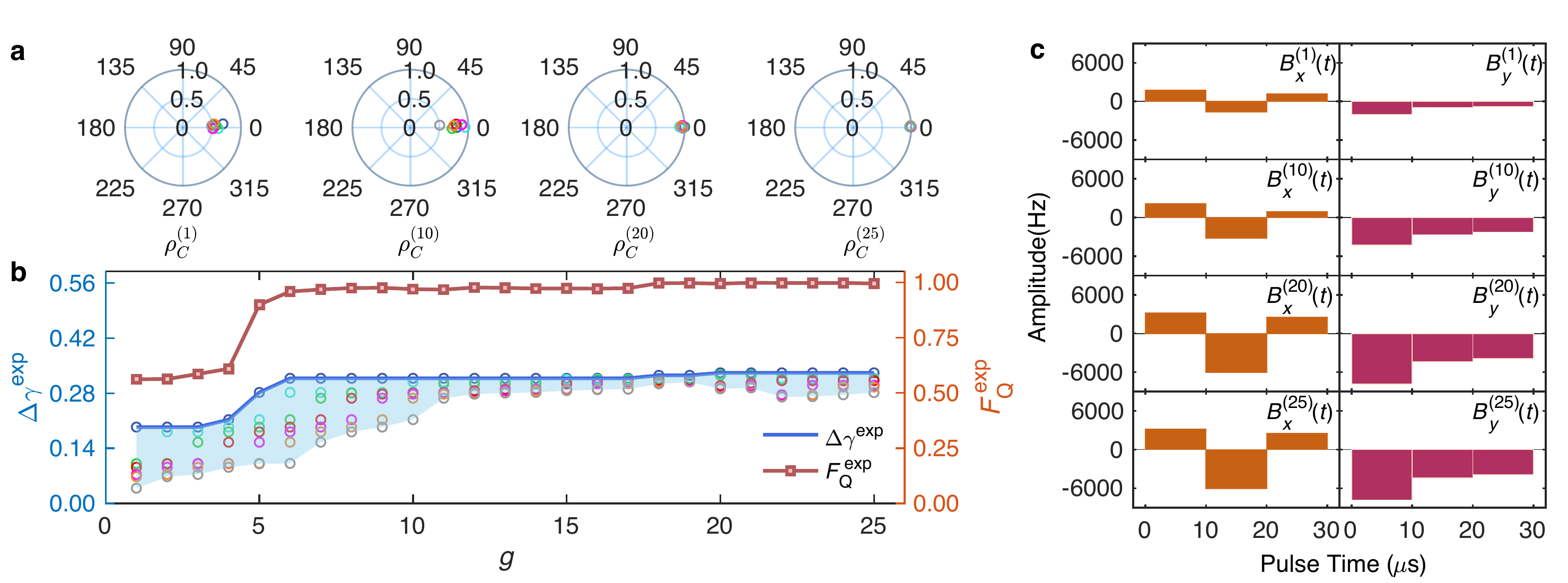}
  \caption{\textbf{Closed-loop learning on the NMR Computer}. In the experimental implementation, each control sequence consisted of three pulses of duration  $T=30~\rm{\mu s}$, with a stopping criterion of 25 iterations and a population size per iteration of $7$. Each candidate probe state in iteration $g$ is specified by $\rho^{(g)}_C=|\psi\rangle \langle\psi|$ with $|\psi\rangle=\cos(\delta/2)|0\rangle+\sin(\delta/2)e^{\mathrm{i}\varphi}|1\rangle$. In \textbf{a}, we plot the candidate probes discovered during iterations $1$, $10$, $20$ and $25$, as overhead projections on the Bloch sphere. Here $(\varphi,\sin\delta)$ are effectively mapped to polar coordinates -- such that $\sin{\delta}$ becomes the magnitude (displacement of the point from the center), and $\varphi$ is the angle relative to the $x$-axis. The points in each plot are color coded according to their efficacy. The plots then directly depict the convergence of the sequentially discovered candidate probes to the optimal probe (demarcated by $\delta=\pi/2$).
  \textbf{b} plots the purity loss of these probe state (round circles with values given by axis on the left), together with the blue line indicating the bound stipulated by maximal purity loss out of all candidates in each iteration (solid blue line). The red line plots (with values given by axis on the right) the quantum Fisher information achievable by the associated probe state, should it be used to sense $\phi$. These results illustrate that the learning algorithm converges quickly to near optimal values by the $10^{\rm{th}}$ iteration.  Meanwhile \textbf{c} plots the associated control fields in the $x$ and $y$ directions (orange and pink bars) used to general the optimal probe of iterations $1$, $10$, $20$ and $25$.}\label{result}
\end{figure*}

 In our experiment, we set the stopping condition as $g = 25$. Fig. \ref{result}\textbf{b} plots the resulting purity loss of various control sequences in $\mathcal{C}^{(g)}$ for each iteration $g$. Meanwhile Fig. \ref{result}\textbf{c} shows the sliced control sequences along $x$ and $y$ directions for the maximum purity loss in the $1^{\rm{st}}$, $10^{\rm{th}}$, $20^{\rm{th}}$ and $25^{\rm{th}}$ iteration. We see these control sequences quickly converge, and the resulting purity loss becomes almost maximal within 10 iterations.

To verify that optimizing purity loss indeed optimizes the efficacy of the probe, we experimentally extracted the best candidate probe state, $\rho^{(g)}_C$, in each iteration from a full three-qubit state tomography \cite{LJS02} (see Supplementary Note 3). The corresponding quantum Fisher information $\fish^{(g)}$ are obtained in Fig. \ref{result}\textbf{b}, illustrating the increments in efficacy of the probes closely follow that of increments in purity loss. Moreover, the final quantum Fisher information obtained is $0.9967\pm 0.0014$ (statistical results over the last 8 iterations), which is very close to the theoretical maximum of $1$. Finally Fig. \ref{result}\textbf{a} illustrates candidate probes at various iterations, illustrating how our controls quickly converge on engineering probe states that are maximal coherent with respect the computational basis -- the requirement for a probe to be optimal for estimating $\phi$.

\section*{DISCUSSION}

Here, we proposed a quantum enhanced machine learning protocol for synthesizing effective probes for the purposes of quantum metrology. The protocol enables an automated method to discover what control sequences one should apply to many-body quantum system -- in order to steer into a state ideally suited for probing the phase $\phi$ of some unitary process $e^{-\mathrm{i}\mathcal{H}\phi}$. We experimentally realized a proof-of-principle experiment using  a $3$-qubit NMR processor, where the device was able to discover control sequences which prepare probe states whose sensitivity to a desired $\phi$ (as measured by quantum Fisher information) is within 1\% of theoretical optimal values. Our numerics indicate this methodology can remain effective when engineering probes involving a large number of entangled qubits - even when these qubits possess uncontrollable spin-spin interactions.

There are a number of open questions. The first is the issue of noise. One of the benefits of our approach is that it automatically accounts for noise during the control process, and naturally finds the optimal control sequence that accounts for such noise. However, the evaluation of purity loss does require the addition of an extra controlled-SWAP gate, and extra noise introduced at this stage can potentially skew the results. Fortunately, our analysis (see Methods) demonstrates that the protocol is highly resistant to one dominant source of noise in NMR -- dephasing, such that any amount of dephasing noise can be corrected for by repeating our purity estimation protocol by some fixed number that does not scale with the size of the system. Sensitivity to other noise sources needs further investigation, and will likely require full tomographical data of the experimentally realized controlled-SWAP gate to correct.

As with all learning algorithms for solving intractable problems, there are of course caveats. The main one is that our algorithm will not always efficiently find the optimal probe. Like all optimization processes, the Nelder-Mead algorithm can be potentially trapped in local optima. Thus one particular important line of future study would be the performance landscape of purity loss. In instances where this landscape is not ideal, our techniques can support multiple pathways for modification. Nelder-Mead, for example, could be replaced with genetic algorithms, neural networks or other means of machine learning~\cite{ZGS15,BPB16,LYP17}. Meanwhile, there may exist other indicators of efficacy that outperform purity loss in certain settings. Thus our closed-loop architecture could be modified to incorporate many possible alternative means of quantum-aided probe design.

Meanwhile, there will always be an ultimate limit to such learning algorithms. The reason is there is a polynomial equivalence between time-complexity in optimal control and quantum gate complexity \cite{nielsen2006quantum,nielsen2006optimal}. Coupled with knowledge that most quantum circuits cannot be efficiently decomposed into fundamental gates, this means that the optimal probes can easily lie outside the set of states that can be synthesized through a control sequence with free parameters that grow as a polynomial of $N$. In such instances, an ideal solution simply does not exist. However, such situations may in fact represent scenarios where such learning protocols are most useful - for its optimization represents all control sequences that can be implemented in some bounded amount of time. As such the solution presented could be a good approximation for the best quantum probe we can synthesize with limited computation power.

\section*{METHODS}

\textbf{Purity measurement in NMR}\\ 
To establish the purity of $\rho^{(g)}_{\mathrm{avg}}$, we made use of stratified sampling. Let $x_k$ be drawn by the stratified sampling method from the discretized Gaussian distribution with $K$ samples and a variance of $(\Delta x)^2 = 1.0721$.

\begin{figure*}
  \centering
  \includegraphics[width=0.85\textwidth,height=0.62\textwidth]{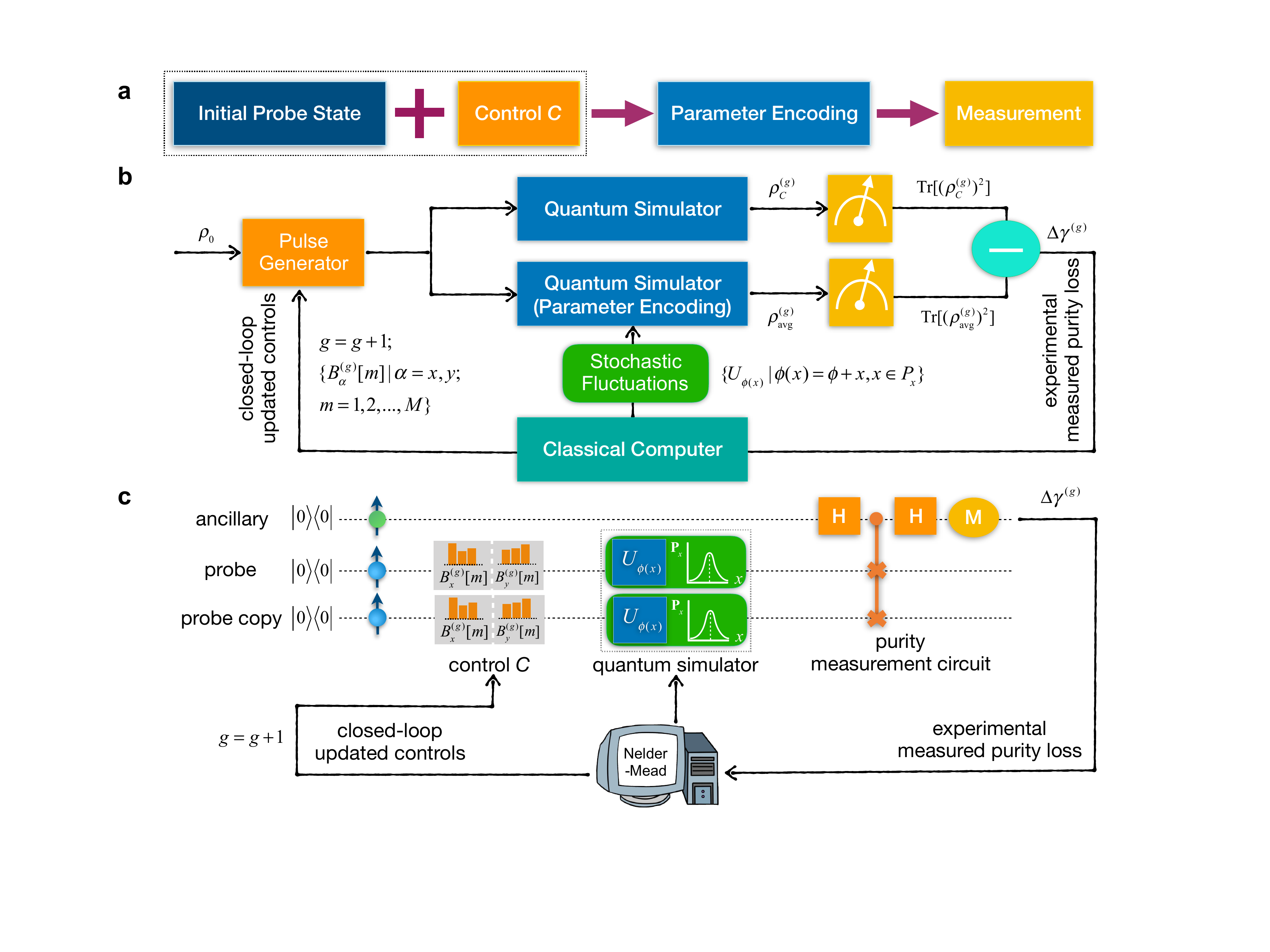}\\
  \caption{\textbf{Schematic diagram of closed-loop learning assisted practical quantum metrology.} \textbf{a}, General procedure of quantum metrology, including probe state preparation (applying controls to the initial probe state to generate a candidate probe $\rho_C$), encoding some parameter $\phi$ by application of $U_\phi$ to $\rho_C$, and measurement read-out. \textbf{b}, A candidate control sequence $C$ is evaluated for efficacy through quantum information processing. This involves using $C$ to prepare copies of the candidate probe state $\rho_C$, half of which are transformed into $\rho_{\mathrm{avg}}$. The purities of $\rho_C$ and $\rho_{\mathrm{avg}}$ are then measured, and their difference - the purity loss - is used as a proxy for efficacy. \textbf{c} illustrates implementation of this process in experiment. The parameter encoding with fluctuations in the dotted box are switched off to determine the purity of $\rho_C$, and on to determine the purity of $\rho_{\mathrm{avg}}$. Their difference is fed into a classical computer running a Nelder-Mead algorithm that generates candidate control sequences for subsequent iterations. In our experiment, this process is automated, such that control fields are tuned automatically at each iteration.}  \label{scheme}
  \end{figure*}

In our experiments, we divided the Gaussian distribution into $K=9$ stratas, such that $x_k \in \{-1.7046, -0.9757, -0.5922, -0.2832, 0, 0.2832, 0.5922, 0.9757,1.7046 \}$, $(\Delta x)^2 = 1.0721$. Let $\rho_{x_j} = U_{\phi({x_j})} \rho^{(g)}_C U_{\phi({x_j})}^{\dag}$ with  $U_{\phi({x_j})} = e^{-\mathrm{i} (\phi + x_j) I_z}$. The purity of the ensemble-averaged state, namely $\rho^{(g)}_{\mathrm{avg}}$, can then be estimated as follows:
\begin{eqnarray}
\text{Tr}[(\rho^{(g)}_{\mathrm{avg}})^2]  &\approx& \text{Tr} [(\frac{1}{K} \sum\limits_{j = 1}^K \rho _{x_j})(\frac{1}{K} \sum\limits_{k = 1}^K \rho _{x_k})] \nonumber \\
&=& \frac{1}{K^2} \sum\limits_{j, k = 1}^K \text{Tr} (\rho _{x_j} \rho _{x_k} ) \\ \nonumber
&=& \frac{1}{K^2}\sum\limits_{j = k}^K {\text{Tr}(\rho _{{x_j}}^2)}  + \frac{2}{K^2}\sum\limits_{j < k}^K {\text{Tr}({\rho _{{x_j}}}{\rho _{{x_k}}})}.
\end{eqnarray}
Hence estimation of the purity $\text{Tr}[(\rho^{(g)}_{\mathrm{avg}})^2]$ was achieved by measuring the purity of each term of $\text{Tr} (\rho _{x_j} \rho _{x_k} )$ using the scheme of Fig. \ref{scheme}\textbf{c}, where qubit $2$ and $3$, respectively, were prepared in $\rho _{x_j}$ and $\rho _{x_k}$. 

\noindent
\textbf{The Nelder-Mead algorithm}\\
The Nelder-Mead algorithm functions by performing a series of geometric transformations on a simplex iteratively to get closer to the optimal control sequence. The simplex is a geometric shape consisting of $n+1$ vertices, and each vertex represents a candidate control sequence $C_{i}$ with $i=1,2,...,n+1$. Here, $n$ should be the product of the directions of the control sequence and and its sliced numbers. Note that $n$ is closely related to the number of vertices. 
Based on the following defined performance function (relate to the efficacy estimator) with respect to each candidate control, namely $f(C_i)=1-\Delta\gamma(\rho_{C_i})$, this algorithm attempts to replace the worst vertex by a new better one according to the geometric transformations reflection, expansion, contraction and shrinkage. Concretely, we describe the procedure of the Nelder-Mead algorithm used in this study.\\
 \textbf{Step 1}: Randomly generate an initial simplex with vertices $\left\{ {{C_1},{C_2}, \cdots ,{C_{n + 1}}} \right\}$ and calculate their performance functions $f_i=f({C}_i)=1-\Delta\gamma(\rho_{C_i})$. The amplitude of $C_i$ in each slice is set in the range $[-1000,1000]$. \\
\textbf{Step 2}: Sort the vertices so that $f({C_1}) \le f({C_2}) \le  \cdots  \le f(C_{n + 1})$, calculate the centroid of the best $n$ points by $\bar{C} = \sum\nolimits_{i = 1}^n {{C_i}}$. \\
\textbf{Step 3}: Calculate the reflected point, ${C_\mathrm r} = \bar {C} + \alpha ({C_{n + 1}} - {C_n})$, evaluate the performance function $f_\mathrm r=f(C_\mathrm r)$, where the reflection factor is set as $\alpha=1$. \\
\textbf{Step 4}: Replace the worst vertex $C_{n+1}$ and its corresponding performance function $f_{n+1}$ by the generated better one according to one of the following conditions:\\
\textbf{(1)} if ${f_1 \le f_\mathrm r < f_n}$, let $C_{n+1}=C_\mathrm r,f_{n+1}=f_\mathrm r$; \\
\textbf{(2)} if ${f_\mathrm r < f_1}$. Calculate the expanded point ${C_\mathrm e} = \bar{C} + \gamma*\alpha ({C_{n + 1}} - {C_n})$, evaluate its performance $f_\mathrm e=f(C_\mathrm e)$, where the expansion factor is set as $\gamma=2$.
\textbf{(2a)} if ${f_\mathrm e < f_\mathrm r}$, let $C_{n+1}=C_\mathrm e,f_{n+1}=f_\mathrm e$;
\textbf{(2b)} if ${f_\mathrm e > f_\mathrm r}$, let $C_{n+1}=C_\mathrm r,f_{n+1}=f_\mathrm r$; \\
\textbf{(3)} if ${f_\mathrm r \geq f_n}$,
\textbf{(3a)} if ${f_n \leq f_\mathrm r < f_{n+1}}$, calculate the outside contracted point, ${C_\mathrm c} = \bar {C} + \beta*\alpha ({C_{n + 1}} - {C_n})$, evaluate the function value $f_\mathrm c = f(C_\mathrm c)$, where the contraction factor is set as $\beta=0.5$. Let $C_{n+1}=C_\mathrm c, f_{n+1}=f_\mathrm c$ when $f_\mathrm c \le f_\mathrm r$, or shrink the simplex ${C_i} = {C_1} + (1 - \delta ){C_i},{f_i} = f({C_i}),i = 2,3, \cdots ,n + 1$ when $f_\mathrm c  > f_\mathrm r$, where $\delta$ is the shrinkage factor and set as 0.5.
\textbf{(3b)} if ${f_\mathrm r \geq f_{n+1}}$, calculate the inside contracted point, ${C_\mathrm c} = \bar {C} - \beta*\alpha ({C_{n + 1}} - {C_n})$, evaluate the function value $f_\mathrm c = f(C_\mathrm c)$, where the contraction factor is set as $\beta=0.5$. Let $C_{n+1}=C_\mathrm c, f_{n+1}=f_\mathrm c$ when $f_\mathrm c \le f_\mathrm r$, or shrink the simplex ${C_i} = {C_1} + (1 - \delta ){C_i},{f_i} = f({C_i}),i = 2,3, \cdots ,n + 1$ when $f_\mathrm c  > f_\mathrm r$, where $\delta$ is the shrinkage factor and set as 0.5.\\
\textbf{Step 5}: Check the stopping conditions, if not satisfied, change the iteration number with $g=g+1$ and continue at Step 2.

\noindent
\textbf{Effects of Decoherence} \\
Here we analyze the effect of decoherence in algorithm. This is because our process for benchmarking the efficacy of a control sequence makes use of the same quantum device that will be used to during the actual metrological process. Specifically, each iteration of the learning algorithm can be casted as the following procedures:
\begin{itemize}
    \item[(A)] Synthesize two copies of candidate probe states $\rho_C$ corresponding to a candidate control sequence $C$. 
    \item[(B)] Synthesize two copies of the state $\rho_{\mathrm{avg}}$, by first preparing a second pair of copies of the candidate probe state $\rho_C$, and then applying the physical encoding process of parameter $\phi$ subject to stochastic fluctuations separately to each copy.
    \item[(C)] Estimate the purity loss due to stochastic fluctuations by experimentally measuring the purity of the resulting states from step (A) and step (B). 
\end{itemize}
We can now consider the impact of docoherence in each of these three steps. The first thing to note is that decoherence in step (A) and (B) respectively represent the intrinsic decoherence of our probe preparation device and that of the physical process it is trying to sense. As such, their inclusion in our learning process is actually desired. That is, as the device that is used to estimate the efficacy of the probes is the device that will eventually be used for metrology; we naturally want all decoherence that within this device to be accounted for while benchmarking the efficacy of candidate control sequences. A similar argument also holds for decoherence when applying the physical process, as this decoherence will also exist during sensing.

Given these considerations, the only undesired decoherence is that which occurs during estimation of purity loss (Step C). This procedure is done via the SWAP test, summarized as follows:
\begin{itemize}
\item[(i)] Take two copies of $\rho_C$, and one ancillary qubit initialized in state $|+\rangle=(|0\rangle+|1\rangle)/\sqrt{2}$.
\item[(ii)] Apply a controlled-SWAP gate to swap the pair of $\rho_C$, add a Hadmard gate to the ancillary qubit, and measure the expectation value $\langle I_z \rangle$ of ancillary control qubit in the $I_z$-basis (see Fig. \ref{scheme}c), to estimate $\mathrm{Tr}(\rho^2_{C})$.
\item[(iii)] Repeat the above procedure for $\rho_{avg}$ to estimate $\mathrm{Tr}(\rho^2_{\mathrm{avg}})$.
\item[(iv)] The difference $\Delta \gamma=\mathrm{Tr}(\rho_C^2)-\mathrm{Tr}(\rho_{\mathrm{avg}}^2)$ is then used to estimate the efficacy of the control sequence $C$.
\end{itemize}
Noise and decoherence during this procedure can affect the accuracy in which we estimate purity loss. In general, its effect is likely non-trivial, and tomography will be needed to work out what  noise introduces to $\langle I_z \rangle$ so that this error can be corrected for.

In the case of NMR, the dominant source arises from dephasing. This dephasing noise can be described by a non-unitary channel $\varepsilon^i(\rho)=(1-p)\rho+4p I_z^i \rho I_z^i$ that acts on each qubit seperately, where $I_z^i$ denotes the angular momentum operator acting on the $i$-th qubit and $p$ is the strength of the dephasing. Following an error analysis similar to that of other NMR experiments that employ controlled gates \cite{PWL10}, we see that this noise does not change the relative order of our purity loss estimates. That is, provided there is a sufficient number of repetitions, our conclusion of which control sequence has greater purity loss between two candidates will not change under dephasing. 

In particular, let $\langle I_z \rangle_p$ denote the expectation value of $I_z$ under dephasing strength $p$, then our measured purity has expectation value $\langle I_z \rangle_p = (1-p)^2\langle I_z \rangle$ with variance bounded above by $1$. To correctly compare two probe states whose purity loss differs by at most $\delta$ requires each purity measurement to have a variance less than $\delta^2/4$ (as differences in purity loss involve four additive purity measurements). This is guaranteed provided we repeat our measurement process of order $\frac{4}{\delta (1-p)^2}$ times -- a overhead of $1/(1-p)^2$ compared to the case where there is no decoherence. Notably this overhead does not scale with $N$, and thus the protocol remains efficient. In our experiment, $p$ is approximately $0.025$, thus we are able to discern rank control sequences whose purity loss differ by more than $0.045$.

\noindent
\textbf{DATA AVAILABILITY}\\
The datasets generated during and/or analysed during the current study are available from the corresponding author on reasonable request.

\noindent
\textbf{ACKNOWLEDGEMENTS} \\
  This work was supported by National Key Research and Development Program of China (Grant No. 2018YFA0306600), National Natural Science Foundation of China (Grants Nos. 11661161018 and 11927811), Anhui Initiative in Quantum Information Technologies (Grant No. AHY050000), the National Research Foundation (NRF) Singapore, under its NRFF Fellow programme (Award No. NRF-NRFF2016-02), the Singapore Ministry of Education Tier 1 Grant 2017-T1-002-043 and 2019-T1-002-015, the NRF-ANR Grant NRF2017-NRF-ANR004 VanQuTe, and the FQXi large grant: FQXi- RFP-1809 the role of quantum effects in simplifying adaptive agents, and FQXi-RFP- IPW-1903 are quantum agents more energetically efficient at making predictions. Any opinions, findings and conclusions or recommendations expressed in this material are those of the author(s) and do not reflect the views of National Research Foundation, Singapore.

\noindent
\textbf{COMPETING INTERESTS}\\ 
 The authors declare that there are no competing interests.

  \noindent
\textbf{AUTHOR CONTRIBUTIONS}\\
 X.P. initiated the project. X.P., J.T. and M.G. conceived the basic procedure. X.P. and X.Y designed the experimental protocol. X.Y. carried out the experiment and analysed the data. All authors contributed to discussing the results and writing the manuscript. 
 
   \noindent
\textbf{ADDITIONAL INFORMATION}\\ 
 Supplementary Information is available for this paper.
 
 \noindent
\textbf{CORRESPONDENCE}\\ 
Correspondence and requests for materials
should be addressed to M.G.~(email: gumile@ntu.edu.sg) or X.P.~(email: xhpeng@ustc.edu.cn).

\noindent
\textbf{REFERENCES}

\providecommand{\noopsort}[1]{}\providecommand{\singleletter}[1]{#1}%
\begin{thebibliography}{10}
\expandafter\ifx\csname url\endcsname\relax
  \def\url#1{\texttt{#1}}\fi
\expandafter\ifx\csname urlprefix\endcsname\relax\def\urlprefix{URL }\fi
\providecommand{\bibinfo}[2]{#2}
\providecommand{\eprint}[2][]{\url{#2}}

\bibitem{GV06}
\bibinfo{author}{Giovannetti, V.}, \bibinfo{author}{Lloyd, S.} \&
  \bibinfo{author}{Maccone, L.}
\newblock \bibinfo{title}{Quantum metrology}.
\newblock \emph{\bibinfo{journal}{Phys. Rev. Lett.}}
  \textbf{\bibinfo{volume}{96}}, \bibinfo{pages}{010401}
  (\bibinfo{year}{2006}).
\newblock
  \urlprefix\url{https://link.aps.org/doi/10.1103/PhysRevLett.96.010401}.

\bibitem{GI11}
\bibinfo{author}{Giovannetti, V.}, \bibinfo{author}{Lloyd, S.} \&
  \bibinfo{author}{Maccone, L.}
\newblock \bibinfo{title}{Advances in quantum metrology}.
\newblock \emph{\bibinfo{journal}{Nature photonics}}
  \textbf{\bibinfo{volume}{5}}, \bibinfo{pages}{222--229}
  (\bibinfo{year}{2011}).
\newblock \urlprefix\url{https://doi.org/10.1038/nphoton.2011.35}.

\bibitem{GT14}
\bibinfo{author}{T{\'{o}}th, G.} \& \bibinfo{author}{Apellaniz, I.}
\newblock \bibinfo{title}{Quantum metrology from a quantum information science
  perspective}.
\newblock \emph{\bibinfo{journal}{Journal of Physics A: Mathematical and
  Theoretical}} \textbf{\bibinfo{volume}{47}}, \bibinfo{pages}{424006}
  (\bibinfo{year}{2014}).
\newblock \urlprefix\url{https://doi.org/10.1088/1751-8113/47/42/424006}.

\bibitem{TQH13}
\bibinfo{author}{Tan, Q.-S.}, \bibinfo{author}{Huang, Y.},
  \bibinfo{author}{Yin, X.}, \bibinfo{author}{Kuang, L.-M.} \&
  \bibinfo{author}{Wang, X.}
\newblock \bibinfo{title}{Enhancement of parameter-estimation precision in
  noisy systems by dynamical decoupling pulses}.
\newblock \emph{\bibinfo{journal}{Phys. Rev. A}} \textbf{\bibinfo{volume}{87}},
  \bibinfo{pages}{032102} (\bibinfo{year}{2013}).
\newblock \urlprefix\url{https://link.aps.org/doi/10.1103/PhysRevA.87.032102}.

\bibitem{LJL15}
\bibinfo{author}{Lang, J.~E.}, \bibinfo{author}{Liu, R.~B.} \&
  \bibinfo{author}{Monteiro, T.~S.}
\newblock \bibinfo{title}{Dynamical-decoupling-based quantum sensing: Floquet
  spectroscopy}.
\newblock \emph{\bibinfo{journal}{Phys. Rev. X}} \textbf{\bibinfo{volume}{5}},
  \bibinfo{pages}{041016} (\bibinfo{year}{2015}).
\newblock \urlprefix\url{https://link.aps.org/doi/10.1103/PhysRevX.5.041016}.

\bibitem{SPS16}
\bibinfo{author}{Sekatski, P.}, \bibinfo{author}{Skotiniotis, M.},
  \bibinfo{author}{Ko{\l}ody{\'n}ski, J.} \& \bibinfo{author}{D{\"u}r, W.}
\newblock \bibinfo{title}{Quantum metrology with full and fast quantum control}
  \eprint{arXiv:1603.08944}.

\bibitem{PMW16}
\bibinfo{author}{Sekatski, P.}, \bibinfo{author}{Skotiniotis, M.} \&
  \bibinfo{author}{Dür, W.}
\newblock \bibinfo{title}{Dynamical decoupling leads to improved scaling in
  noisy quantum metrology}.
\newblock \emph{\bibinfo{journal}{New Journal of Physics}}
  \textbf{\bibinfo{volume}{18}}, \bibinfo{pages}{073034}
  (\bibinfo{year}{2016}).
\newblock \urlprefix\url{https://doi.org/10.1088/1367-2630/18/7/073034}.

\bibitem{WMF14}
\bibinfo{author}{D\"ur, W.}, \bibinfo{author}{Skotiniotis, M.},
  \bibinfo{author}{Fr\"owis, F.} \& \bibinfo{author}{Kraus, B.}
\newblock \bibinfo{title}{Improved quantum metrology using quantum error
  correction}.
\newblock \emph{\bibinfo{journal}{Phys. Rev. Lett.}}
  \textbf{\bibinfo{volume}{112}}, \bibinfo{pages}{080801}
  (\bibinfo{year}{2014}).
\newblock
  \urlprefix\url{https://link.aps.org/doi/10.1103/PhysRevLett.112.080801}.

\bibitem{KLS14}
\bibinfo{author}{Kessler, E.~M.}, \bibinfo{author}{Lovchinsky, I.},
  \bibinfo{author}{Sushkov, A.~O.} \& \bibinfo{author}{Lukin, M.~D.}
\newblock \bibinfo{title}{Quantum error correction for metrology}.
\newblock \emph{\bibinfo{journal}{Phys. Rev. Lett.}}
  \textbf{\bibinfo{volume}{112}}, \bibinfo{pages}{150802}
  (\bibinfo{year}{2014}).
\newblock
  \urlprefix\url{https://link.aps.org/doi/10.1103/PhysRevLett.112.150802}.

\bibitem{YF15}
\bibinfo{author}{Yuan, H.} \& \bibinfo{author}{Fung, C.-H.~F.}
\newblock \bibinfo{title}{Optimal feedback scheme and universal time scaling
  for hamiltonian parameter estimation}.
\newblock \emph{\bibinfo{journal}{Phys. Rev. Lett.}}
  \textbf{\bibinfo{volume}{115}}, \bibinfo{pages}{110401}
  (\bibinfo{year}{2015}).
\newblock
  \urlprefix\url{https://link.aps.org/doi/10.1103/PhysRevLett.115.110401}.

\bibitem{YHD16}
\bibinfo{author}{Yuan, H.}
\newblock \bibinfo{title}{Sequential feedback scheme outperforms the parallel
  scheme for hamiltonian parameter estimation}.
\newblock \emph{\bibinfo{journal}{Phys. Rev. Lett.}}
  \textbf{\bibinfo{volume}{117}}, \bibinfo{pages}{160801}
  (\bibinfo{year}{2016}).
\newblock
  \urlprefix\url{https://link.aps.org/doi/10.1103/PhysRevLett.117.160801}.

\bibitem{LJY17}
\bibinfo{author}{Liu, J.} \& \bibinfo{author}{Yuan, H.}
\newblock \bibinfo{title}{Quantum parameter estimation with optimal control}.
\newblock \emph{\bibinfo{journal}{Phys. Rev. A}} \textbf{\bibinfo{volume}{96}},
  \bibinfo{pages}{012117} (\bibinfo{year}{2017}).
\newblock \urlprefix\url{https://link.aps.org/doi/10.1103/PhysRevA.96.012117}.

\bibitem{MMJ04}
\bibinfo{author}{Mitchell, M.~W.}, \bibinfo{author}{Lundeen, J.~S.} \&
  \bibinfo{author}{Steinberg, A.~M.}
\newblock \bibinfo{title}{Super-resolving phase measurements with a multiphoton
  entangled state}.
\newblock \emph{\bibinfo{journal}{Nature}} \textbf{\bibinfo{volume}{429}},
  \bibinfo{pages}{161} (\bibinfo{year}{2004}).
\newblock \urlprefix\url{https://doi.org/10.1038/nature02493}.

\bibitem{LKS05}
\bibinfo{author}{Leibfried, D.} \emph{et~al.}
\newblock \bibinfo{title}{Creation of a six-atom 'schr{\"o}dinger cat'state}.
\newblock \emph{\bibinfo{journal}{Nature}} \textbf{\bibinfo{volume}{438}},
  \bibinfo{pages}{639} (\bibinfo{year}{2005}).
\newblock \urlprefix\url{https://doi.org/10.1038/nature04251}.

\bibitem{RPP07}
\bibinfo{author}{Resch, K.~J.} \emph{et~al.}
\newblock \bibinfo{title}{Time-reversal and super-resolving phase
  measurements}.
\newblock \emph{\bibinfo{journal}{Phys. Rev. Lett.}}
  \textbf{\bibinfo{volume}{98}}, \bibinfo{pages}{223601}
  (\bibinfo{year}{2007}).
\newblock
  \urlprefix\url{https://link.aps.org/doi/10.1103/PhysRevLett.98.223601}.

\bibitem{JJM11}
\bibinfo{author}{Joo, J.}, \bibinfo{author}{Munro, W.~J.} \&
  \bibinfo{author}{Spiller, T.~P.}
\newblock \bibinfo{title}{Quantum metrology with entangled coherent states}.
\newblock \emph{\bibinfo{journal}{Phys. Rev. Lett.}}
  \textbf{\bibinfo{volume}{107}}, \bibinfo{pages}{083601}
  (\bibinfo{year}{2011}).
\newblock
  \urlprefix\url{https://link.aps.org/doi/10.1103/PhysRevLett.107.083601}.

\bibitem{ACDM02}
\bibinfo{author}{Ekert, A.~K.} \emph{et~al.}
\newblock \bibinfo{title}{Direct estimations of linear and nonlinear
  functionals of a quantum state}.
\newblock \emph{\bibinfo{journal}{Phys. Rev. Lett.}}
  \textbf{\bibinfo{volume}{88}}, \bibinfo{pages}{217901}
  (\bibinfo{year}{2002}).
\newblock
  \urlprefix\url{https://link.aps.org/doi/10.1103/PhysRevLett.88.217901}.

\bibitem{EDR11}
\bibinfo{author}{Escher, B.~M.}, \bibinfo{author}{de~Matos~Filho, R.~L.} \&
  \bibinfo{author}{Davidovich, L.}
\newblock \bibinfo{title}{General framework for estimating the ultimate
  precision limit in noisy quantum-enhanced metrology}.
\newblock \emph{\bibinfo{journal}{Nature Physics}}
  \textbf{\bibinfo{volume}{7}}, \bibinfo{pages}{406} (\bibinfo{year}{2011}).
\newblock \urlprefix\url{https://doi.org/10.1038/nphys1958}.

\bibitem{EBMM11}
\bibinfo{author}{Escher, B.~M.}, \bibinfo{author}{de~Matos~Filho, R.~L.} \&
  \bibinfo{author}{Davidovich, L.}
\newblock \bibinfo{title}{Quantum metrology for noisy systems}.
\newblock \emph{\bibinfo{journal}{Brazilian Journal of Physics}}
  \textbf{\bibinfo{volume}{41}}, \bibinfo{pages}{229--247}
  (\bibinfo{year}{2011}).
\newblock \urlprefix\url{https://doi.org/10.1007/s13538-011-0037-y}.

\bibitem{HLW12}
\bibinfo{author}{Hyllus, P.} \emph{et~al.}
\newblock \bibinfo{title}{Fisher information and multiparticle entanglement}.
\newblock \emph{\bibinfo{journal}{Phys. Rev. A}} \textbf{\bibinfo{volume}{85}},
  \bibinfo{pages}{022321} (\bibinfo{year}{2012}).
\newblock \urlprefix\url{https://link.aps.org/doi/10.1103/PhysRevA.85.022321}.

\bibitem{LJS02}
\bibinfo{author}{Lee, J.-S.}
\newblock \bibinfo{title}{The quantum state tomography on an nmr system}.
\newblock \emph{\bibinfo{journal}{Physics Letters A}}
  \textbf{\bibinfo{volume}{305}}, \bibinfo{pages}{349--353}
  (\bibinfo{year}{2002}).
\newblock \urlprefix\url{https://doi.org/10.1016/S0375-9601(02)01479-2}.

\bibitem{ZGS15}
\bibinfo{author}{Zahedinejad, E.}, \bibinfo{author}{Ghosh, J.} \&
  \bibinfo{author}{Sanders, B.~C.}
\newblock \bibinfo{title}{High-fidelity single-shot toffoli gate via quantum
  control}.
\newblock \emph{\bibinfo{journal}{Phys. Rev. Lett.}}
  \textbf{\bibinfo{volume}{114}}, \bibinfo{pages}{200502}
  (\bibinfo{year}{2015}).
\newblock
  \urlprefix\url{https://link.aps.org/doi/10.1103/PhysRevLett.114.200502}.

\bibitem{BPB16}
\bibinfo{author}{Banchi, L.}, \bibinfo{author}{Pancotti, N.} \&
  \bibinfo{author}{Bose, S.}
\newblock \bibinfo{title}{Quantum gate learning in qubit networks: Toffoli gate
  without time-dependent control}.
\newblock \emph{\bibinfo{journal}{npj Quantum Information}}
  \textbf{\bibinfo{volume}{2}}, \bibinfo{pages}{16019} (\bibinfo{year}{2016}).
\newblock \urlprefix\url{https://doi.org/10.1038/npjqi.2016.19}.

\bibitem{LYP17}
\bibinfo{author}{Li, J.}, \bibinfo{author}{Yang, X.}, \bibinfo{author}{Peng,
  X.} \& \bibinfo{author}{Sun, C.-P.}
\newblock \bibinfo{title}{Hybrid quantum-classical approach to quantum optimal
  control}.
\newblock \emph{\bibinfo{journal}{Phys. Rev. Lett.}}
  \textbf{\bibinfo{volume}{118}}, \bibinfo{pages}{150503}
  (\bibinfo{year}{2017}).
\newblock
  \urlprefix\url{https://link.aps.org/doi/10.1103/PhysRevLett.118.150503}.

\end{thebibliography}


\begin{thebibliography}{10}
\expandafter\ifx\csname url\endcsname\relax
  \def\url#1{\texttt{#1}}\fi
\expandafter\ifx\csname urlprefix\endcsname\relax\def\urlprefix{URL }\fi
\providecommand{\bibinfo}[2]{#2}
\providecommand{\eprint}[2][]{\url{#2}}

\bibitem{GV06}
\bibinfo{author}{Giovannetti, V.}, \bibinfo{author}{Lloyd, S.} \&
  \bibinfo{author}{Maccone, L.}
\newblock \bibinfo{title}{Quantum metrology}.
\newblock \emph{\bibinfo{journal}{Phys. Rev. Lett.}}
  \textbf{\bibinfo{volume}{96}}, \bibinfo{pages}{010401}
  (\bibinfo{year}{2006}).

\bibitem{GI11}
\bibinfo{author}{Giovannetti, V.}, \bibinfo{author}{Lloyd, S.} \&
  \bibinfo{author}{Maccone, L.}
\newblock \bibinfo{title}{Advances in quantum metrology}.
\newblock \emph{\bibinfo{journal}{Nat. Photonics}}
  \textbf{\bibinfo{volume}{5}}, \bibinfo{pages}{222--229}
  (\bibinfo{year}{2011}).

\bibitem{GT14}
\bibinfo{author}{T{\'{o}}th, G.} \& \bibinfo{author}{Apellaniz, I.}
\newblock \bibinfo{title}{Quantum metrology from a quantum information science
  perspective}.
\newblock \emph{\bibinfo{journal}{J. Phys. A: Math. Theor.}}
  \textbf{\bibinfo{volume}{47}}, \bibinfo{pages}{424006}
  (\bibinfo{year}{2014}).

\bibitem{KLS14}
\bibinfo{author}{Kessler, E.~M.}, \bibinfo{author}{Lovchinsky, I.},
  \bibinfo{author}{Sushkov, A.~O.} \& \bibinfo{author}{Lukin, M.~D.}
\newblock \bibinfo{title}{Quantum error correction for metrology}.
\newblock \emph{\bibinfo{journal}{Phys. Rev. Lett.}}
  \textbf{\bibinfo{volume}{112}}, \bibinfo{pages}{150802}
  (\bibinfo{year}{2014}).

\bibitem{LJL15}
\bibinfo{author}{Lang, J.~E.}, \bibinfo{author}{Liu, R.~B.} \&
  \bibinfo{author}{Monteiro, T.~S.}
\newblock \bibinfo{title}{Dynamical-decoupling-based quantum sensing: Floquet
  spectroscopy}.
\newblock \emph{\bibinfo{journal}{Phys. Rev. X}} \textbf{\bibinfo{volume}{5}},
  \bibinfo{pages}{041016} (\bibinfo{year}{2015}).

\bibitem{YF15}
\bibinfo{author}{Yuan, H.} \& \bibinfo{author}{Fung, C.-H.~F.}
\newblock \bibinfo{title}{Optimal feedback scheme and universal time scaling
  for hamiltonian parameter estimation}.
\newblock \emph{\bibinfo{journal}{Phys. Rev. Lett.}}
  \textbf{\bibinfo{volume}{115}}, \bibinfo{pages}{110401}
  (\bibinfo{year}{2015}).

\bibitem{YHD16}
\bibinfo{author}{Yuan, H.}
\newblock \bibinfo{title}{Sequential feedback scheme outperforms the parallel
  scheme for hamiltonian parameter estimation}.
\newblock \emph{\bibinfo{journal}{Phys. Rev. Lett.}}
  \textbf{\bibinfo{volume}{117}}, \bibinfo{pages}{160801}
  (\bibinfo{year}{2016}).

\bibitem{HWJ19}
\bibinfo{author}{Hou, Z.} \emph{et~al.}
\newblock \bibinfo{title}{Control-enhanced sequential scheme for general
  quantum parameter estimation at the heisenberg limit}.
\newblock \emph{\bibinfo{journal}{Phys. Rev. Lett.}}
  \textbf{\bibinfo{volume}{123}}, \bibinfo{pages}{040501}
  (\bibinfo{year}{2019}).

\bibitem{OLO12}
\bibinfo{author}{Okamoto, R.} \emph{et~al.}
\newblock \bibinfo{title}{Experimental demonstration of adaptive quantum state
  estimation}.
\newblock \emph{\bibinfo{journal}{Phys. Rev. Lett.}}
  \textbf{\bibinfo{volume}{109}}, \bibinfo{pages}{130404}
  (\bibinfo{year}{2012}).

\bibitem{HBD07}
\bibinfo{author}{Higgins, B.~L.}, \bibinfo{author}{Berry, D.~W.},
  \bibinfo{author}{Bartlett, S.~D.}, \bibinfo{author}{Wiseman, H.~M.} \&
  \bibinfo{author}{Pryde, G.~J.}
\newblock \bibinfo{title}{Entanglement-free heisenberg-limited phase
  estimation}.
\newblock \emph{\bibinfo{journal}{Nature}} \textbf{\bibinfo{volume}{450}},
  \bibinfo{pages}{393} (\bibinfo{year}{2007}).

\bibitem{BGN15}
\bibinfo{author}{Berni, A.~A.} \emph{et~al.}
\newblock \bibinfo{title}{Ab initio quantum-enhanced optical phase estimation
  using real-time feedback control}.
\newblock \emph{\bibinfo{journal}{Nat. Photonics}}
  \textbf{\bibinfo{volume}{9}}, \bibinfo{pages}{577} (\bibinfo{year}{2015}).

\bibitem{BBD16}
\bibinfo{author}{Bonato, C.} \emph{et~al.}
\newblock \bibinfo{title}{Optimized quantum sensing with a single electron spin
  using real-time adaptive measurements}.
\newblock \emph{\bibinfo{journal}{Nat. Nanotechnol.}}
  \textbf{\bibinfo{volume}{11}}, \bibinfo{pages}{247} (\bibinfo{year}{2016}).

\bibitem{HS10}
\bibinfo{author}{Hentschel, A.} \& \bibinfo{author}{Sanders, B.~C.}
\newblock \bibinfo{title}{Machine learning for precise quantum measurement}.
\newblock \emph{\bibinfo{journal}{Phys. Rev. Lett.}}
  \textbf{\bibinfo{volume}{104}}, \bibinfo{pages}{063603}
  (\bibinfo{year}{2010}).

\bibitem{PGS17}
\bibinfo{author}{Paesani, S.} \emph{et~al.}
\newblock \bibinfo{title}{Experimental bayesian quantum phase estimation on a
  silicon photonic chip}.
\newblock \emph{\bibinfo{journal}{Phys. Rev. Lett.}}
  \textbf{\bibinfo{volume}{118}}, \bibinfo{pages}{100503}
  (\bibinfo{year}{2017}).

\bibitem{PJ17}
\bibinfo{author}{Pang, S.} \& \bibinfo{author}{Jordan, A.~N.}
\newblock \bibinfo{title}{Optimal adaptive control for quantum metrology with
  time-dependent hamiltonians}.
\newblock \emph{\bibinfo{journal}{Nat. Commun.}} \textbf{\bibinfo{volume}{8}},
  \bibinfo{pages}{14695} (\bibinfo{year}{2017}).

\bibitem{PCL12}
\bibinfo{author}{Pan, J.-W.} \emph{et~al.}
\newblock \bibinfo{title}{Multiphoton entanglement and interferometry}.
\newblock \emph{\bibinfo{journal}{Rev. Mod. Phys.}}
  \textbf{\bibinfo{volume}{84}}, \bibinfo{pages}{777} (\bibinfo{year}{2012}).

\bibitem{JKF09}
\bibinfo{author}{Jones, J.~A.} \emph{et~al.}
\newblock \bibinfo{title}{Magnetic field sensing beyond the standard quantum
  limit using 10-spin noon states}.
\newblock \emph{\bibinfo{journal}{Science}} \textbf{\bibinfo{volume}{324}},
  \bibinfo{pages}{1166--1168} (\bibinfo{year}{2009}).

\bibitem{SJK10}
\bibinfo{author}{Simmons, S.}, \bibinfo{author}{Jones, J.~A.},
  \bibinfo{author}{Karlen, S.~D.}, \bibinfo{author}{Ardavan, A.} \&
  \bibinfo{author}{Morton, J. J.~L.}
\newblock \bibinfo{title}{Magnetic field sensors using 13-spin cat states}.
\newblock \emph{\bibinfo{journal}{Phys. Rev. A}} \textbf{\bibinfo{volume}{82}},
  \bibinfo{pages}{022330} (\bibinfo{year}{2010}).

\bibitem{MMJ04}
\bibinfo{author}{Mitchell, M.~W.}, \bibinfo{author}{Lundeen, J.~S.} \&
  \bibinfo{author}{Steinberg, A.~M.}
\newblock \bibinfo{title}{Super-resolving phase measurements with a multiphoton
  entangled state}.
\newblock \emph{\bibinfo{journal}{Nature}} \textbf{\bibinfo{volume}{429}},
  \bibinfo{pages}{161} (\bibinfo{year}{2004}).

\bibitem{RPP07}
\bibinfo{author}{Resch, K.~J.} \emph{et~al.}
\newblock \bibinfo{title}{Time-reversal and super-resolving phase
  measurements}.
\newblock \emph{\bibinfo{journal}{Phys. Rev. Lett.}}
  \textbf{\bibinfo{volume}{98}}, \bibinfo{pages}{223601}
  (\bibinfo{year}{2007}).

\bibitem{JJM11}
\bibinfo{author}{Joo, J.}, \bibinfo{author}{Munro, W.~J.} \&
  \bibinfo{author}{Spiller, T.~P.}
\newblock \bibinfo{title}{Quantum metrology with entangled coherent states}.
\newblock \emph{\bibinfo{journal}{Phys. Rev. Lett.}}
  \textbf{\bibinfo{volume}{107}}, \bibinfo{pages}{083601}
  (\bibinfo{year}{2011}).

\bibitem{EDR11}
\bibinfo{author}{Escher, B.~M.}, \bibinfo{author}{de~Matos~Filho, R.~L.} \&
  \bibinfo{author}{Davidovich, L.}
\newblock \bibinfo{title}{General framework for estimating the ultimate
  precision limit in noisy quantum-enhanced metrology}.
\newblock \emph{\bibinfo{journal}{Nat. Phys.}} \textbf{\bibinfo{volume}{7}},
  \bibinfo{pages}{406} (\bibinfo{year}{2011}).

\bibitem{EBMM11}
\bibinfo{author}{Escher, B.~M.}, \bibinfo{author}{de~Matos~Filho, R.~L.} \&
  \bibinfo{author}{Davidovich, L.}
\newblock \bibinfo{title}{Quantum metrology for noisy systems}.
\newblock \emph{\bibinfo{journal}{Braz. J. Phys.}}
  \textbf{\bibinfo{volume}{41}}, \bibinfo{pages}{229--247}
  (\bibinfo{year}{2011}).

\bibitem{WCL16}
\bibinfo{author}{Wang, X.-L.} \emph{et~al.}
\newblock \bibinfo{title}{Experimental ten-photon entanglement}.
\newblock \emph{\bibinfo{journal}{Phys. Rev. Lett.}}
  \textbf{\bibinfo{volume}{117}}, \bibinfo{pages}{210502}
  (\bibinfo{year}{2016}).

\bibitem{LKS05}
\bibinfo{author}{Leibfried, D.} \emph{et~al.}
\newblock \bibinfo{title}{Creation of a six-atom schr{\"o}dinger cat'state}.
\newblock \emph{\bibinfo{journal}{Nature}} \textbf{\bibinfo{volume}{438}},
  \bibinfo{pages}{639} (\bibinfo{year}{2005}).

\bibitem{JD13}
\bibinfo{author}{Jarzyna, M.} \&
  \bibinfo{author}{Demkowicz-Dobrza\ifmmode~\acute{n}\else \'{n}\fi{}ski, R.}
\newblock \bibinfo{title}{Matrix product states for quantum metrology}.
\newblock \emph{\bibinfo{journal}{Phys. Rev. Lett.}}
  \textbf{\bibinfo{volume}{110}}, \bibinfo{pages}{240405}
  (\bibinfo{year}{2013}).

\bibitem{MLJM16}
\bibinfo{author}{Modi, K.}, \bibinfo{author}{C{\'e}leri, L.~C.},
  \bibinfo{author}{Thompson, J.} \& \bibinfo{author}{Gu, M.}
\newblock \bibinfo{title}{Fragile states are better for quantum metrology}
  (\bibinfo{year}{2016}).
\newblock \eprint{arXiv:1608.01443}.

\bibitem{ACDM02}
\bibinfo{author}{Ekert, A.~K.} \emph{et~al.}
\newblock \bibinfo{title}{Direct estimations of linear and nonlinear
  functionals of a quantum state}.
\newblock \emph{\bibinfo{journal}{Phys. Rev. Lett.}}
  \textbf{\bibinfo{volume}{88}}, \bibinfo{pages}{217901}
  (\bibinfo{year}{2002}).

\bibitem{LTT00}
\bibinfo{author}{Lewis, R.~M.}, \bibinfo{author}{Torczon, V.} \&
  \bibinfo{author}{Trosset, M.~W.}
\newblock \bibinfo{title}{Direct search methods: then and now}.
\newblock \emph{\bibinfo{journal}{J. Comput. Appl. Math.}}
  \textbf{\bibinfo{volume}{124}}, \bibinfo{pages}{191--207}
  (\bibinfo{year}{2000}).

\bibitem{ES15}
\bibinfo{author}{Eiben, A.~E.} \& \bibinfo{author}{Smith, J.}
\newblock \bibinfo{title}{From evolutionary computation to the evolution of
  things}.
\newblock \emph{\bibinfo{journal}{Nature}} \textbf{\bibinfo{volume}{521}},
  \bibinfo{pages}{476} (\bibinfo{year}{2015}).

\bibitem{NM65}
\bibinfo{author}{Nelder, J.~A.} \& \bibinfo{author}{Mead, R.}
\newblock \bibinfo{title}{A simplex method for function minimization}.
\newblock \emph{\bibinfo{journal}{The computer journal}}
  \textbf{\bibinfo{volume}{7}}, \bibinfo{pages}{308--313}
  (\bibinfo{year}{1965}).

\bibitem{DS13}
\bibinfo{author}{Devoret, M.~H.} \& \bibinfo{author}{Schoelkopf, R.~J.}
\newblock \bibinfo{title}{Superconducting circuits for quantum information: an
  outlook}.
\newblock \emph{\bibinfo{journal}{Science}} \textbf{\bibinfo{volume}{339}},
  \bibinfo{pages}{1169--1174} (\bibinfo{year}{2013}).

\bibitem{BCM19}
\bibinfo{author}{Bruzewicz, C.~D.}, \bibinfo{author}{Chiaverini, J.},
  \bibinfo{author}{McConnell, R.} \& \bibinfo{author}{Sage, J.~M.}
\newblock \bibinfo{title}{Trapped-ion quantum computing: Progress and
  challenges}.
\newblock \emph{\bibinfo{journal}{Applied Physics Reviews}}
  \textbf{\bibinfo{volume}{6}}, \bibinfo{pages}{021314} (\bibinfo{year}{2019}).

\bibitem{OJF09}
\bibinfo{author}{O'brien, J.~L.}, \bibinfo{author}{Furusawa, A.} \&
  \bibinfo{author}{Vu{\v{c}}kovi{\'c}, J.}
\newblock \bibinfo{title}{Photonic quantum technologies}.
\newblock \emph{\bibinfo{journal}{Nat. Photonics}}
  \textbf{\bibinfo{volume}{3}}, \bibinfo{pages}{687} (\bibinfo{year}{2009}).

\bibitem{HLW12}
\bibinfo{author}{Hyllus, P.} \emph{et~al.}
\newblock \bibinfo{title}{Fisher information and multiparticle entanglement}.
\newblock \emph{\bibinfo{journal}{Phys. Rev. A}} \textbf{\bibinfo{volume}{85}},
  \bibinfo{pages}{022321} (\bibinfo{year}{2012}).

\bibitem{LJS02}
\bibinfo{author}{Lee, J.-S.}
\newblock \bibinfo{title}{The quantum state tomography on an nmr system}.
\newblock \emph{\bibinfo{journal}{Phys. Lett. A}}
  \textbf{\bibinfo{volume}{305}}, \bibinfo{pages}{349--353}
  (\bibinfo{year}{2002}).

\bibitem{ZGS15}
\bibinfo{author}{Zahedinejad, E.}, \bibinfo{author}{Ghosh, J.} \&
  \bibinfo{author}{Sanders, B.~C.}
\newblock \bibinfo{title}{High-fidelity single-shot toffoli gate via quantum
  control}.
\newblock \emph{\bibinfo{journal}{Phys. Rev. Lett.}}
  \textbf{\bibinfo{volume}{114}}, \bibinfo{pages}{200502}
  (\bibinfo{year}{2015}).

\bibitem{BPB16}
\bibinfo{author}{Banchi, L.}, \bibinfo{author}{Pancotti, N.} \&
  \bibinfo{author}{Bose, S.}
\newblock \bibinfo{title}{Quantum gate learning in qubit networks: Toffoli gate
  without time-dependent control}.
\newblock \emph{\bibinfo{journal}{npj Quantum Inf.}}
  \textbf{\bibinfo{volume}{2}}, \bibinfo{pages}{16019} (\bibinfo{year}{2016}).

\bibitem{LYP17}
\bibinfo{author}{Li, J.}, \bibinfo{author}{Yang, X.}, \bibinfo{author}{Peng,
  X.} \& \bibinfo{author}{Sun, C.-P.}
\newblock \bibinfo{title}{Hybrid quantum-classical approach to quantum optimal
  control}.
\newblock \emph{\bibinfo{journal}{Phys. Rev. Lett.}}
  \textbf{\bibinfo{volume}{118}}, \bibinfo{pages}{150503}
  (\bibinfo{year}{2017}).

\bibitem{nielsen2006quantum}
\bibinfo{author}{Nielsen, M.~A.}, \bibinfo{author}{Dowling, M.~R.},
  \bibinfo{author}{Gu, M.} \& \bibinfo{author}{Doherty, A.~C.}
\newblock \bibinfo{title}{Quantum computation as geometry}.
\newblock \emph{\bibinfo{journal}{Science}} \textbf{\bibinfo{volume}{311}},
  \bibinfo{pages}{1133--1135} (\bibinfo{year}{2006}).

\bibitem{nielsen2006optimal}
\bibinfo{author}{Nielsen, M.~A.}, \bibinfo{author}{Dowling, M.~R.},
  \bibinfo{author}{Gu, M.} \& \bibinfo{author}{Doherty, A.~C.}
\newblock \bibinfo{title}{Optimal control, geometry, and quantum computing}.
\newblock \emph{\bibinfo{journal}{Phys. Rev. A}} \textbf{\bibinfo{volume}{73}},
  \bibinfo{pages}{062323} (\bibinfo{year}{2006}).

\bibitem{PWL10}
\bibinfo{author}{Peng, X.}, \bibinfo{author}{Wu, S.}, \bibinfo{author}{Li, J.},
  \bibinfo{author}{Suter, D.} \& \bibinfo{author}{Du, J.}
\newblock \bibinfo{title}{Observation of the ground-state geometric phase in a
  heisenberg $xy$ model}.
\newblock \emph{\bibinfo{journal}{Phys. Rev. Lett.}}
  \textbf{\bibinfo{volume}{105}}, \bibinfo{pages}{240405}
  (\bibinfo{year}{2010}).

\end{thebibliography}

\providecommand{\noopsort}[1]{}\providecommand{\singleletter}[1]{#1}%

 
\end{document}


\title{Supplemental Material to ``Probe optimization for quantum metrology via closed-loop learning control''}
\author{Xiaodong Yang}
\affiliation{CAS Key Laboratory of Microscale Magnetic Resonance and Department of Modern Physics, University of Science and Technology of China, Hefei, Anhui 230026, China}

\author{Jayne Thompson}
\affiliation{Centre for Quantum Technologies, National University of Singapore, 3 Science Drive 2, Singapore 117543}

\author{Ze Wu}
\affiliation{CAS Key Laboratory of Microscale Magnetic Resonance and Department of Modern Physics, University of Science and Technology of China, Hefei, Anhui 230026, China}

\author{Mile Gu}
\email{cqtmileg@nus.edu.sg}
\affiliation{School of Physical and Mathematical Sciences, Nanyang Technological University, Singapore and Centre for Quantum Technologies, National University of Singapore, Singapore}
\affiliation{Complexity Institute, Nanyang Technological University,
Singapore and Centre for Quantum Technologies, National University of Singapore, Singapore}
\affiliation{Centre for Quantum Technologies, National University of Singapore, 3 Science Drive 2, Singapore 117543}
\author{Xinhua Peng}
\email{xhpeng@ustc.edu.cn}
\affiliation{CAS Key Laboratory of Microscale Magnetic Resonance and Department of Modern Physics, University of Science and Technology of China, Hefei, Anhui 230026, China}
\affiliation{Hefei National Laboratory for Physical Sciences at the Microscale, University of Science and Technology of China, Hefei 230026, China}
\affiliation{Synergetic Innovation Centre of Quantum Information $\&$ Quantum Physics, University of Science and Technology of China, Hefei, Anhui 230026, China}

\author{Jiangfeng Du}
\affiliation{CAS Key Laboratory of Microscale Magnetic Resonance and Department of Modern Physics, University of Science and Technology of China, Hefei, Anhui 230026, China}
\affiliation{Hefei National Laboratory for Physical Sciences at the Microscale, University of Science and Technology of China, Hefei 230026, China}\affiliation{Synergetic Innovation Centre of Quantum Information $\&$ Quantum Physics, University of Science and Technology of China, Hefei, Anhui 230026, China}

\date{\today}
\maketitle

\renewcommand{\thesubsection}{\arabic{subsection}}
\renewcommand{\theequation}{S\arabic{equation}}
\renewcommand{\thefigure}{S\arabic{figure}}
\renewcommand{\thetable}{S\arabic{table}}
\renewcommand{\citenumfont}[1]{S#1} 

\noindent
\textbf{\large Note S1. Numerical simulation details when sensing with spin chains}

\bigskip
\noindent
To show the advantages of the proposed protocol, we consider estimating the phase $\phi$ on $N$-qubit spin chains evolving nearest-neighbor Ising couplings $\mathcal{H}_{S}=2 \pi J \sum_{i=2}^{N} I_{z}^{i-1} I_{z}^{i}$ , as described in the main text. Local control fields along $x$ and $y$ directions of each qubit are applied to search the optimal probe, therefore each slice of the control sequence can be expressed as ${U_m} = {e^{ - \mathrm{i}\Delta t_m 2\pi \{ J\sum_{i = 2}^N {I_z^{i - 1}I_z^i}  + \sum_{i = 1}^N {(B_x^i[m]I_x^i}  + B_y^i[m]I_y^i)\} }}$, where $I_{x,y,z}^i$ are the spin angular momentum operators of the $i^{\mathrm{th}}$ qubit and $J$ is a settled coupling constant. 
In order to search the optimal controls more conveniently and easily, we transform the above evolution operator $U_m$ to another usual form ${U^{\prime}_m} = {e^{ - \mathrm{i}\Delta \tau 2\pi \{ B^{\prime}_J[m]\sum_{i = 2}^N {I_z^{i - 1}I_z^i}  + B_x^{\prime}[m]\sum_{i = 1}^N {I_x^i} + B_y^{\prime}[m]\sum_{i = 1}^N {I_y^i} \} }}.$ To make the above two forms equivalent, we need to meet the conditions $\Delta t[m] J = \Delta \tau {B'_J}[m],~\Delta t[m] B_x^i[m] = \Delta \tau {B'_x}[m],~\Delta t[m] B_y^i[m] = \Delta \tau {B'_y}[m]$,  where we set $J$ and $\Delta \tau$ as constant values during the numerical optimization process. Then the conversion rules can be written as $\Delta t[m] = {{{{B'_J}}[m]\Delta \tau }}/{J},B_x^i[m] = {{{B'_x}[m]J}}/{{{B'_J}[m]}},B_y^i[m] = {{{B'_y}[m]J}}/{{{B'_J}[m]}}$. 

The numerical searching process starts from an easily prepared initial state $|{\Psi _\mathrm i} \rangle  = | 0 \rangle ^{\otimes N}$, and stops when sufficient iterations are performed, such that the final probe converges to some optimal state $|{\Psi _\mathrm f}\rangle$ through optimizing the control parameters$C=(B_{x}^{'}[m], B_{y}^{'}[m], B_{J}^{'}[m])$(or equivalently $C = (B^i_{x}[m], B^i_{y}[m], \Delta t[m])$). Theory tell us the optimal states for this case can be expressed as the form of NOON states, i.e., $|{\Psi _\mathrm t}\rangle=(| 0 \rangle ^{\otimes N}+ e^{\mathrm{i}\theta}| 1 \rangle ^{\otimes N})/\sqrt{2}$. We have listed their overlaps in the main text, and here we list the non-zero matrix elements of $\rho_\mathrm f=|\Psi _\mathrm f\rangle \langle \Psi _\mathrm f|$, i.e., $\rho_\mathrm f^{11},\rho_\mathrm f^{1n},\rho_\mathrm f^{n1},\rho_\mathrm f^{nn}$, in Table. \ref{sim}, where $n=2^N$ is the matrix dimension. 

\begin{table}[h]
\begin{center}
\begin{tabular}{|p{1cm}<{\centering}|p{2cm}<{\centering}|p{3cm}<{\centering}|p{3cm}<{\centering}|p{2cm}<{\centering}|}
\hline
  $N$ & $\rho_\mathrm f^{11}$ & $\rho_\mathrm f^{1n}$ & $\rho_\mathrm f^{n1}$ & $\rho_\mathrm f^{nn}$ \\ \hline
1 & 0.5000 & 0.3131+0.3899i & 0.3131-0.3899i & 0.5000 \\ \hline
2 & 0.5000 & 0.1409-0.4797i & 0.1409+0.4797i & 0.5000 \\ \hline
3 & 0.5018 & 0.4058-0.2921i & 0.4058+0.2921i & 0.4981 \\ \hline
4 & 0.4849 & 0.3739-0.3021i & 0.3739+0.3021i & 0.4765 \\ \hline
5 & 0.4857 & -0.4846+0.0253i & -0.4846-0.0253i & 0.4847 \\ \hline
6 & 0.4959 & -0.2668-0.4070i & -0.2668+0.4070i & 0.4776 \\ \hline
7 & 0.4870 & -0.0213+0.4748i & -0.0213-0.4748i & 0.4637 \\ 
\hline
\end{tabular}
\end{center}
\caption{\label{sim}\textbf{Numerical simulation results when sensing with spin chains.} The non-zero elements of the optimal probe state $\rho_\mathrm f=|\Psi _\mathrm f\rangle \langle \Psi _\mathrm f|$ searched numerically, and the optimal probe state is engineered by the optimal controls $U_C$ from the initial probe state $|{\Psi _\mathrm i} \rangle$, i.e., $|{\Psi_\mathrm f} \rangle= U_C|{\Psi _\mathrm i} \rangle$.}
\end{table}

\bigskip
\noindent
\textbf{\large Note S2. Simulation method for the Gaussian fluctuation}

\bigskip
\noindent
In our protocol, the to-be-estimated parameter $\phi$ is supposed to undergo some stochastic fluctuations. Though the types of these fluctuations can be quite diverse, providing that they are sharply distributed around $\phi$ \cite{M16}, a common and reasonable expectation should be Gaussian distribution. Here, we establish a systematic method to simulate this kind of Gaussian fluctuation.

\begin{figure}[htbp!]
  \centering
\includegraphics[width=0.75\textwidth,height=0.55\textwidth]{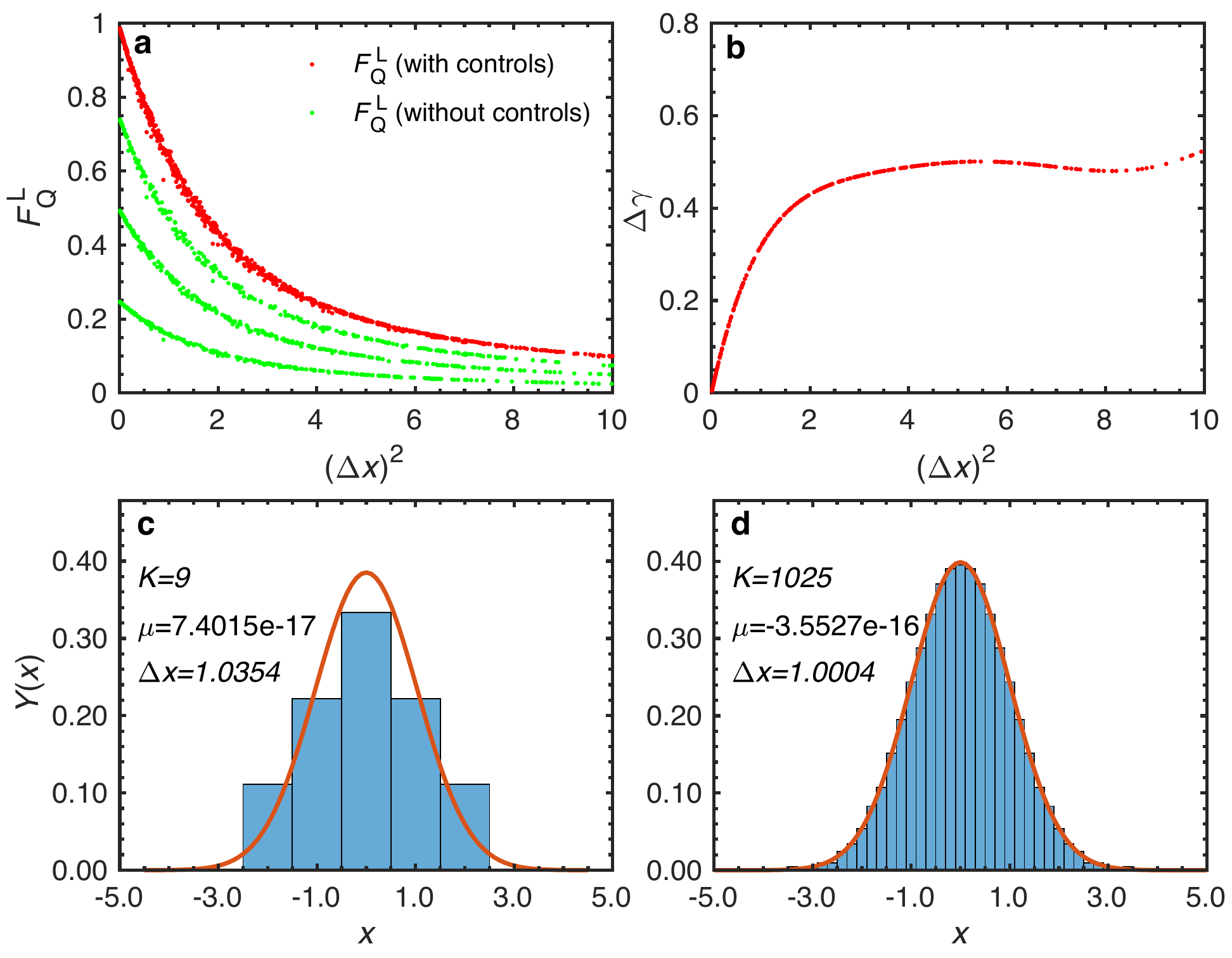}
  \caption{\textbf{Choosing the parameters for simulating the Gaussian fluctuation.} \textbf{a}, $F_\mathrm{Q}^\mathrm{L}$ for several probe states (green/red dots for the cases without/with controls) when the parameter $\phi$ undergoes Gaussian fluctuation of different strength $\Delta x$. \textbf{b} shows the corresponding maximum purity loss with respect to different strength $\Delta x$ on $\phi$. \textbf{c} and \textbf{d} plot the frequency histogram and its fitted probability density function when the sample size is $K=9$ and $K=1025$, respectively. The mean and the variance of the target Gaussian distribution are chosen as $\mu=0,(\Delta x)^2=1$.}\label{correctcheck}
 \end{figure}

The density function for Gaussian distribution can be expressed as $Y(x)=\frac{1}{\sqrt{2\pi}\Delta x} e^{-(x-\mu)^2/2(\Delta x)^2}$, the corresponding probability function is $G(x)=\frac{1}{2}[1+\text{erf}(\frac{x-\mu}{2(\Delta x)^2})]$, and the inverse probability function is $x=G^{-1}(z)=\sqrt{2}\Delta x*\text{erfinv}(2z-1)+\mu$, where $\text{erf}(\cdot)$ is the error function and $\text{erfinv}(\cdot)$ is the inverse error function. In order to conveniently choose $K$ samples from the distribution, we use the stratified sampling method. First, we divide the total probability into $K$ equal ranges, thus the range bounds are $t_k=G^{-1}(k/K),k=1,2,...,K-1$. Afterwards, we select one sample per range using the following way: $x_1=\frac{1}{K}\int_{\rm{-inf}}^{t_1} xY(x)\mathrm{d}x,x_K=\frac{1}{K}\int_{t_{K-1}}^{\rm{inf}} xY(x)\mathrm{d}x,x_j=\frac{1}{K}\int_{t_{j-1}}^{t_{j}} xY(x)\mathrm{d}x,j=2,3,...,K-1$. Finally, $K$ samples are successfully selected which can fit a very accurate target Gaussian distribution.

With the above Gaussian distribution simulation method in hand, we now choose the parameters $\Delta x$ and $K$. Consider single-particle probe states here, we show $F_\mathrm{Q}^\mathrm{L}$ for several probe states (just take three examples) when the parameter $\phi$ undergoes Gaussian fluctuation of different strength $\Delta x$ in Fig. \ref{correctcheck}\textbf{a}. One can easily find that the controls will engineer different initial pure probe states to the same optimal one (maybe with some global phase). Therefore, the red dots show the biggest $F_\mathrm{Q}^\mathrm{L}$ we can get for the chosen $\Delta x$ no matter what the initial probe state is. What's more, we should balance the biggest $F_\mathrm{Q}^\mathrm{L}$ with the direct observation value, i.e., the purity loss $\Delta\gamma=F_\mathrm{Q}^\mathrm{L}(\Delta x)^2/2$, so we plot the corresponding maximum $\Delta\gamma$ with respect to $(\Delta x)^2$ in Fig. \ref{correctcheck}\textbf{b}. Both of $F_\mathrm{Q}^\mathrm{L}$ and $\Delta\gamma$ should be big enough to get reliable and observable results, thus we choose $(\Delta x)^2=1$ in our experiments. On one hand, as we analyzed in the Methods, suppose we select $K$ samples from the Gaussian distribution, the total number of the experiments required to measure $\text{Tr}(\rho_{\mathrm{avg}}^2)$ is $\text{C}(K,2) + K =K(K-1)/2+K$. This indicates that we can't choose too many samples as the complexity increases quadratically. On the other hand, the sample size $K$ should be large enough to fit a reasonable Gaussian distribution while meeting the precision requirements. Using the sampling method described above, we set $\mu=0,(\Delta x)^2=1$ and vary the sample size in a very broad range. Finally, in consideration of the experimental time and the precision, we choose $K=9$ to implement our demonstrative experiments. We show two cases $K=9$ and $K=1025$ in Fig. \ref{correctcheck}\textbf{c},\textbf{d}.

\bigskip
\noindent
\textbf{\large Note S3. NMR techniques for realizing the proposed protocol}

\bigskip
\noindent
\textbf{System initialization.}
The internal Hamiltonian of our 3-qubit NMR processor can be expressed as ${\mathcal{H}_S} =  \sum\nolimits_{i = 1}^3 {\Delta \omega _0^iI_z^i}  + 2\pi\sum\nolimits_{i < j}^3 {{J_{ij}}I_z^i} I_z^j$, where $\Delta \omega_0^i$ represents the offset of the $i$-th spin in the rotating frame, $J_{ij}$ is the $J$-coupling strength between the $i$-th and $j$-th spin, and $I_z^i$ is the spin angular momentum operator of $i$-th spin. Starting from the thermal equilibrium state, we used line-selective method \cite{PZF01} to prepare a pseudo-pure state ${\rho _{\mathrm{pps}}} = \frac{{1 - \varepsilon }}{8}I + \varepsilon \left| {000} \right\rangle \left\langle {000} \right|$, where $\varepsilon \approx 10^{-5}$ represents the thermal polarization of this three-qubit system. As the identity matrix having no physical observable effects, this led to the system initial state $|000\rangle$, by standard state tomography \cite{LJS02}, we confirmed the result with the fidelity being $0.992$.

\begin{figure*}
  \centering
\includegraphics[width=0.9\textwidth,height=0.48\textwidth]{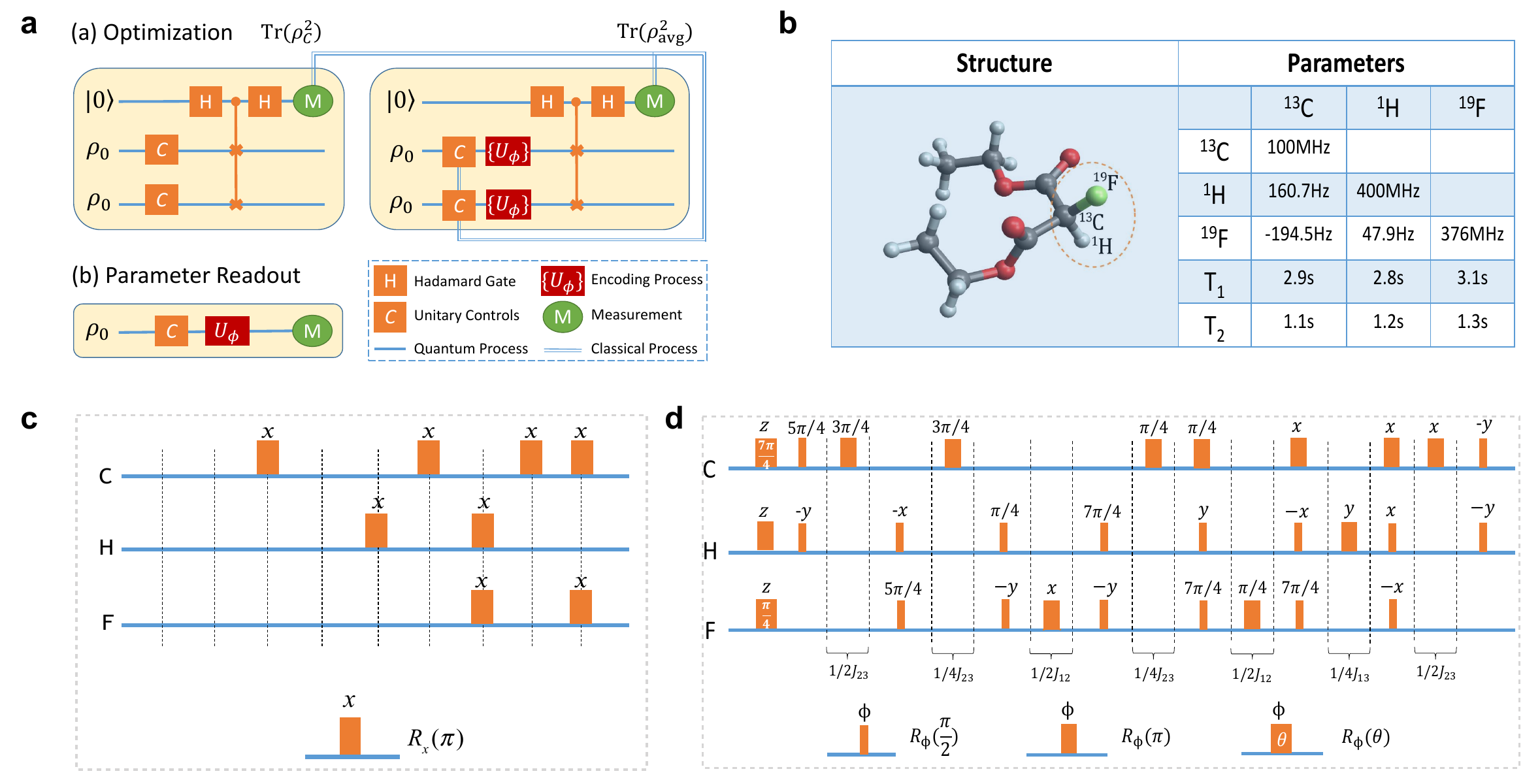}\\
  \caption{\textbf{Experimental circuits, pulse sequences and sample information.}\textbf{a}, The explicit quantum circuits of procedures (a) optimal controls learning, (b) parameter readout. \textbf{b}, Molecular structure and Hamiltonian parameters of the sample $^{13}\text{C}$-unlabeled diethyl-fluoromalonate used in experiments. \textbf{c}, The decoupling pulse sequences for eliminating all the $J$-couplings and the precession of spin 1, leaving out the precessions of spin 2 and 3 to carry out the encoding process. The time intervals between each two nearby dash lines are the same. \textbf{d}, The pulse sequence of two Hardamard gates placed at each side of the controlled-SWAP gate, resulting in the purity measurement circuit, i.e. $\text{H} \text{C}_{\text{swap}} \text{H}$.} \label{experiment}
\end{figure*}

\bigskip
\noindent
\textbf{Decouple unwanted interactions.}
The parameter encoding operator $U_\phi=e^{-\mathrm{i} I_z \phi}$ was operated on the probe spins. As we used a copy of the probe and an ancillary, the total evolution operator should be $I \otimes e^{-\mathrm{i} \phi I_z} \otimes e^{-\mathrm{i} \phi I_z}$. To realize the same evolution on the last two coupled spins, we used the decoupling techniques \cite{DI00} to cancel the effects of the $J$-coupling evolution. Based on the system Hamiltonian ${\mathcal{H}_S}$,  the explicit pulse sequences to achieve this are shown in Fig. \ref{experiment}\textbf{c}, where the offset of these three spins are set as $\Delta \omega_0^1=0,\Delta \omega_0^2=\Delta \omega_0^3=50*2\pi~\rm{rad/s}$. Thus, with some definite encoding time $T_\mathrm e$, the left free evolution of the spins 2 and 3 would induce an encoding parameter $\phi=2\pi\Delta \omega_0^2 T_\mathrm e=2\pi\Delta \omega_0^3 T_\mathrm e$. For the settled parameter $\phi=\pi/3$, $T_\mathrm e=0.003333~\rm{s}$.

\bigskip
\noindent
\textbf{Construct pulse sequences for circuits.}
We demonstrate the explicit quantum circuit of our proposed control sequence learning protocol and the parameter readout procedure in Fig. \ref{experiment}\textbf{a}. To realize these quantum circuits, we need to decompose every quantum gate into a sequence of basic operations that can be realized directly, which is closely related to the specific quantum architecture we are using. Any quantum circuit in NMR system can be decomposed into single-qubit rotations and free evolution of two-spin $I_zI_z$ couplings. Based on the molecular structure and Hamiltonian parameters of the sample $^{13}\text{C}$-unlabeled diethyl-fluoromalonate we used in the experiments, which is listed in Fig. \ref{experiment}\textbf{b},  the pulse sequence including the controlled-SWAP gate and two Hadamard gates, namely the purity measurement circuit, is shown in Fig. \ref{experiment}\textbf{d}. 

\begin{figure*}
  \centering
  \includegraphics[width=0.75\textwidth,height=0.5\textwidth]{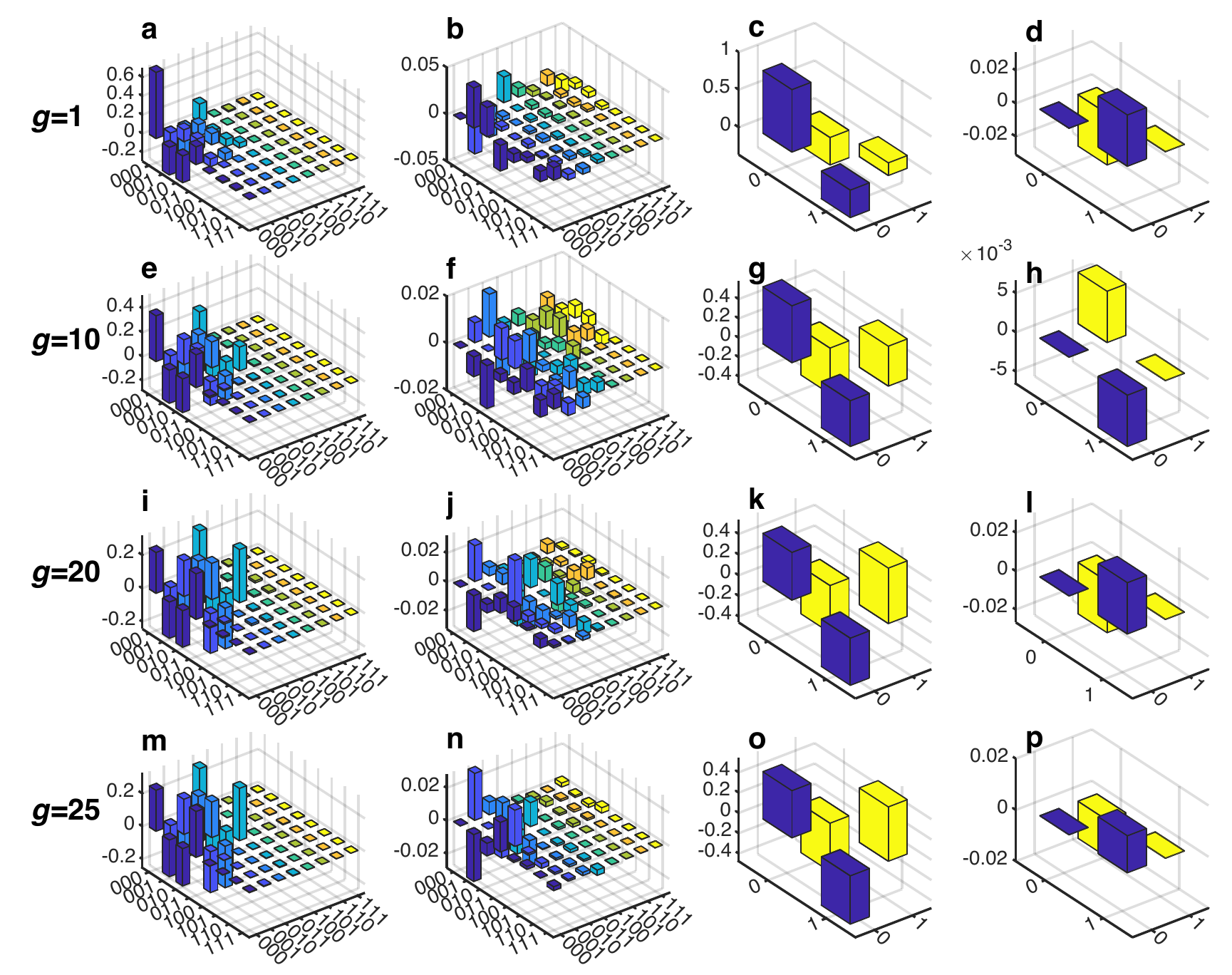}\\
  \caption{\textbf{The full state tomography results for the entire three-qubit system state and the extracted single-qubit probe state.} These probes are the optimal states discovered in the $1^{\mathrm{st}}, 10^{\mathrm{th}}, 20^{\mathrm{th}}$ and $25^{\mathrm{th}}$ iteration. In each row, from left to right, the figures are the real part of the total three-qubit system state, the imaginary of the total three-qubit system state, the real part of the single-qubit probe state and the imaginary of the single-qubit probe state, respectively.  }\label{tomo1}
  \end{figure*}

\bigskip
\noindent
\textbf{State tomography for measuring $F_\mathrm Q$.}
As stated above, the quantum fisher information $F_\mathrm Q(\rho^{(g)}_C)$ of the best candidate probe state in each iteration was obtain by extracting the probe state $\rho^{(g)}_C$ from a full three-qubit state tomography (use partial trace to obtain the density matrix of the probe state). Here, we show the three-qubit full tomography and the reduced single-qubit tomography results in the $1^{\mathrm{st}}$, $10^{\mathrm{th}}$, $20^{\mathrm{th}}$ and $25^{\mathrm{th}}$ iteration in Fig. \ref{tomo1}. In each row of this figure, from left to right, the subfigures are the real part of the total three-qubit probe system state, the imaginary of the total three-qubit probe system state, the real part of the single-qubit probe state and the imaginary of the single-qubit probe state, respectively.

\bigskip
\noindent
\textbf{\large Note S4. Optimal controls learning -- another independent trial}\label{Another}
\begin{figure*}[htbp!]
  \centering
  \includegraphics[width=0.9\textwidth,height=0.33\textwidth]{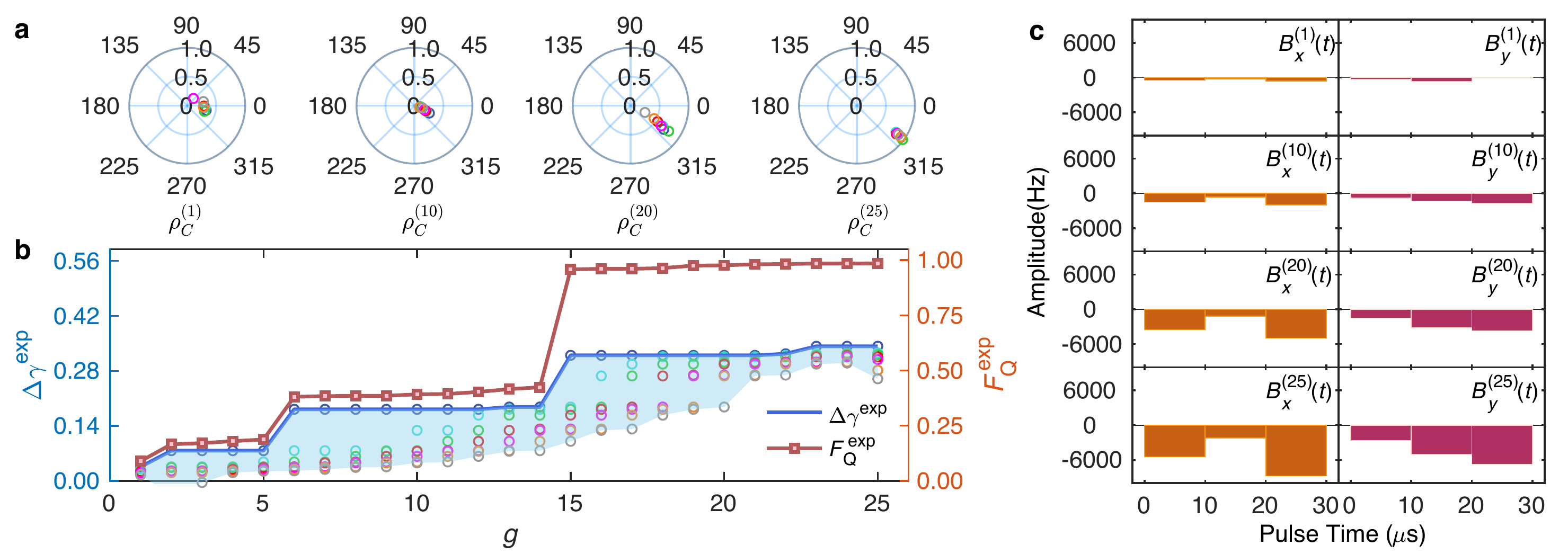}
  \caption{\textbf{Closed-loop learning on the NMR Computer}. In the experimental implementation, each control sequence consisted of three pulses of duration  $T=30~\rm{\mu s}$, with a stopping criterion of 25 iterations and a population size per iteration of $7$. Each candidate probe state in iteration $g$ is specified by $\rho^{(g)}_C=|\psi\rangle \langle\psi|$ with $|\psi\rangle=\cos(\delta/2)|0\rangle+\sin(\delta/2)e^{\mathrm{i}\varphi}|1\rangle$. In \textbf{a} we plot the candidate probes discovered during iterations $1$, $10$, $20$ and $25$, as overhead projections on the Bloch sphere. Here $(\varphi,\sin(\delta))$ are effectively mapped to polar coordinates -- such that $\sin{\delta}$ becomes the magnitude (displacement of the point from the center), and $\varphi$ is the angle relative to the $x$-axis. The points in each plot are color coded according to their efficacy. The plots then directly depict the convergence of the sequentially discovered candidate probes to the optimal probe (demarcated by $\delta=\pi/2$).
  \textbf{b} plots the purity loss of these probe state (round circles with values given by axis on the left), together with the blue line indicating the bound stipulated by maximal purity loss out of all candidates in each iteration (solid blue line). The red line plots (with values given by axis on the right) the quantum Fisher information achievable by the associated probe state, should it be used to sense $\phi$. These results illustrate that the learning algorithm converges quickly to near optimal values by the $10^{\mathrm{th}}$ iteration.  Meanwhile \textbf{c} plots the associated control fields in the $x$ and $y$ directions (orange and pink bars) used to general the optimal probe of iterations $1$, $10$, $20$ and $25$.}\label{result2}
\end{figure*}

\bigskip
\noindent
To further verify our experimental results, we ran the whole controls learning experiments again. Similarly, we set the stopping condition as $g = 25$. Fig. \ref{result2}\textbf{b} plots the resulting purity loss of various control sequences in $\mathcal{C}^{(g)}$ for each iteration $g$. Meanwhile Fig. \ref{result2}\textbf{c} shows the sliced control sequences along $x$ and $y$ directions for the maximum purity loss in the $1^{\mathrm{st}}$, $10^{\mathrm{th}}$, $20^{\mathrm{th}}$ and $25^{\mathrm{th}}$ iteration. We see the these control sequences quickly converge, and that the resulting purity loss becomes almost maximal within 15 iterations.

To verify that optimizing purity loss indeed optimizes the efficacy of the probe, we experimentally extracted the best candidate probe state, $\rho^{(g)}_C$, in each iteration from a full three-qubit state tomography \cite{LJS02}. The corresponding quantum Fisher information $F_{\mathrm Q}^{(g)}$ are obtained in Fig. \ref{result2}\textbf{b}, illustrating the increases in efficacy of the probes closely follow that of increases in purity loss. Moreover, the final Fisher information obtained is $0.9836\pm 0.0017$ (statistical results over the last 5 iterations), which is very close to the theoretical maximum of $1$. Finally Fig. \ref{result2}\textbf{a} illustrates candidate probes at various iterations, illustrating that how our controls quickly convey on engineering probe states that are maximal coherent with respect the computational basis -- the requirement for a probe to be optimal for estimating $\phi$.

\bigskip
\noindent
\textbf{\large Note S5. The procedure of reading the encoded phase $\phi$ out}

\bigskip
\noindent
After finding the optimal control pulses $C_{\mathrm{opt}}$, we could continue to finish the metrology task of estimating $\phi=\pi/3$ on a single probe spin governed by $e^{-\mathrm{i} I_z \phi}$. This procedure was conducted by firstly preparing the total system state as $\left| 0 \rangle \langle 0 \right| \otimes {\left| 0 \rangle \langle 0 \right| } \otimes {\left| 0 \rangle \langle 0 \right| }$. The optimal control $C_{\mathrm{opt}}$ searched in the optimization part was then applied to the probe spin 2. Thereafter, we achieved the parameter encoding process on probe spin 2 by setting $\Delta \omega_0^2=50~\rm{Hz}$ and varying the encoding time $T_\mathrm e$ from $0$ to $0.02~\rm{s}$. Finally, a Hadamard gate was followed to measure on the computational basis $\left| 0 \right\rangle$ with decoupling other two spins. The measurement results formed a curve looks like $f(\omega)=\frac{1}{2} + { \tilde {a_2}}\cos (\tilde\omega t ) - {\tilde {a_3}}\sin (\tilde \omega t )+h$. A fitted $\tilde \omega$  could be gotten to calculate the concerned parameter, i.e. $\tilde \phi=\tilde \omega T_\mathrm e$, where $T_\mathrm e$ should be $0.003333~\rm{s}$ for $\phi=\pi/3$. For comparison, we performed a similar experiment when controls learning were not considered, but the probe spin was prepared at $\rho_1$ (for another trial in Note S4, we marked as $\rho_2$) which was induced by the searched controls at $1^{\mathrm{st}}$ iteration. Here, in Table. \ref{fit} we list the fitting results of the readout curves. As analysed, the fitting curves read $f(\tilde \omega)=\frac{1}{2} + \tilde {a_2} \cos (\tilde{\omega} t ) - \tilde {a_3}\sin (\tilde {\omega} t )+h$.

\begin{figure*}
  \centering
  \includegraphics[width=0.7\textwidth,height=0.28\textwidth]{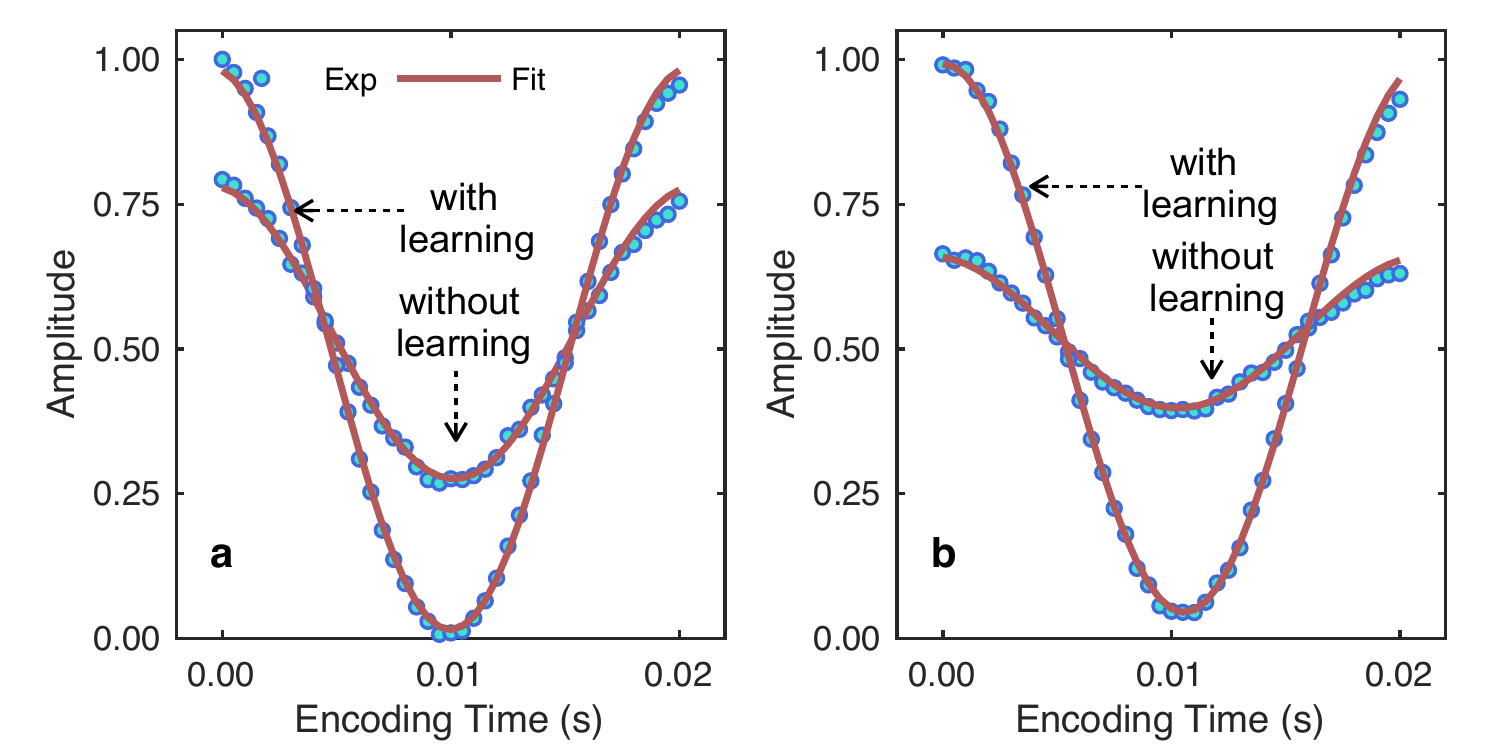}\\
  \caption{\textbf{The measured and fitted results for the encoding parameter under controls learning or not}. The probe state is $\rho_1$ (in the main text) for \textbf{a} and $\rho_2$ (in Note S4) for \textbf{b}. We achieved the parameter encoding process by setting $\Delta \omega_0^2=50~\rm{Hz}$ and varying the encoding time $T_\mathrm e$ from $0$ to $0.02~\rm{s}$. }\label{readout}
\end{figure*}

The experimental results of the parameter readout process are shown in Fig. \ref{readout}. We used the least-squares method to fit the curve and estimate the phase. In addition, we used the fringe visibility $v=(A_{\max}-A_{\min})/(A_{\max}+A_{\min})$ to directly estimate the phase uncertainty \cite{P18} in this noisy experiment, where $A_{\max}$ and $A_{\min}$ are the maximum and minimum of the observed amplitude. When the controls learning are not included, for the probe state $\rho_1$, we obtained $\tilde\phi=0.3128\pi$ and $v=0.4945$. The optimal controls improved this result to $\tilde\phi=0.3217\pi$ with $v=0.9863$, as shown in Fig. \ref{readout}\textbf{a}. Thus, we got a phase sensitivity enhancement factor $0.9863/0.4945=1.9945$. For the probe state $\rho_2$, the results are similar, they are $\tilde\phi=0.3065\pi$ with $v=0.2569$ for the case without controls learning and $\tilde\phi=0.3157\pi$ with $v=0.9142$ when controls learning are considered, as shown in Fig. \ref{readout}\textbf{b}. In this case, we got a phase sensitivity enhancement factor $0.9142/0.2569=3.5586$. From these results, we clearly see that the estimated encoding parameter is more closer to the true value ($\pi/3$) and more accurate when optimal controls are included.

\begin{table*}[htbp]
\begin{center}
\begin{tabular}{|p{5cm}<{\centering}|p{1.5cm}<{\centering}|p{1.5cm}<{\centering}|p{1.5cm}<{\centering}|p{1.5cm}<{\centering}|p{1.5cm}<{\centering}|}
\hline
 & $\tilde \omega$ & $\tilde {a_2}$ & $\tilde {a_3}$ & $h$ & resnorm \\
\hline
without controls learning ($\rho_1$) & 46.9313 & 0.2500 & 0.0388& 0.0287 & 0.0028\\
\hline
with controls learning ($\rho_1$) & 48.2638& 0.4836 & 0.0603& 0.0000 & 0.0026\\
\hline
without controls learning ($\rho_2$) & 45.9747 & 0.1295 & 0.0226 & 0.0292 &0.0018\\
\hline
with controls learning ($\rho_2$) & 47.3524 & 0.4738 & 0.0000& 0.0178 & 0.0031\\
\hline
\end{tabular}
\end{center}
\caption{\label{fit} \textbf{The results of the estimated encoding parameter.} The fitting results of the experimental readout curves $f(\tilde \omega)=\frac{1}{2} + \tilde {a_2} \cos (\tilde{\omega} t ) - \tilde {a_3}\sin (\tilde {\omega} t )+h$ in the case with controls learning and without controls learning, where `resnorm' represents the value of the squared 2-norm of the residual at $\tilde{\omega}$ . }
\end{table*}
